\documentclass[a4paper,11pt,oneside]{article}

\usepackage{jcappub}
\usepackage[margin=1in]{geometry}
\usepackage{physics}
\usepackage{enumerate}
\usepackage{empheq}
\usepackage{booktabs}
\usepackage{amsmath,amssymb,amsbsy,amstext, amsthm, simplewick}
\usepackage{hyperref}
\usepackage{graphicx}
\usepackage{amsfonts}
\usepackage{color}
\usepackage{caption}
\usepackage{wasysym}
\usepackage{wrapfig}
\usepackage{multirow}
\usepackage{array}

\usepackage{colortbl}
\definecolor{summersky}{cmyk}{0.71,0.33,0,0.5}
\definecolor{flamingo}{cmyk}{0,0.51,0.71,0.5}
\definecolor{rp}{cmyk}{0.2, 1, 0.6, 0}
\definecolor{pacificblue}{cmyk}{0.95,0.3,0, 0.5}
\definecolor{gray60}{cmyk}{0.4,0.4,0,0.8}
\hypersetup{
    pdftoolbar=true,        
    pdfmenubar=true,        
    pdffitwindow=true,     
    pdfstartview={FitH},    
    pdfnewwindow=true,      
    colorlinks=true,       
    linkcolor=pacificblue,          
    citecolor=flamingo,        
    filecolor=magenta,      
    urlcolor=pacificblue           
}

\newcommand{\ex}[1]{\langle #1 \rangle}
\newcommand{\be}{\begin{eqnarray} }
\newcommand{\ee}{\end{eqnarray} }
\newcommand{\bs}{\begin{split} }
\newcommand{\es}{\end{split} }

\newcommand{\hvec}[1]{\mathbf{\hat{#1}} }

\renewcommand{\v}[1]{\mathbf{#1} }

\newcommand{\then}{\quad \Rightarrow\quad}
\renewcommand{\O}{\mathcal{O}}

\newcommand{\nc}{\newcommand}
\newcommand{\bfx}{\textbf{x}}
\newcommand{\bfk}{\textbf{k}}
\nc{\nn}{\nonumber}
\nc{\eps}{\epsilon}

\newcommand{\gb}{\bar g}

\newcommand{\D}{\partial}

\title{\centering Spatial Curvature at the Sound Horizon}

\author[a]{Guus Avis,}
\author[b]{Sadra Jazayeri,}
\author[b]{Enrico Pajer,}
\author[b]{Jakub Supe\l{} }

\affiliation[a]{Institute for Theoretical Physics and Center for Extreme Matter and Emergent Phenomena,
	Utrecht University, 
	Princetonplein 5, 3584 CC Utrecht, The Netherlands}
\affiliation[b]{Department of Applied Mathematics and Theoretical Physics, Centre for Mathematical Sciences,
University of Cambridge, Wilberforce Road, Cambridge CB3 0WA, UK}
\emailAdd{ep551@cam.ac.uk}
\emailAdd{sj571@cam.ac.uk}
\emailAdd{js2154@cam.ac.uk}

\abstract{\noindent  
The effect of spatial curvature on primordial perturbations is controlled by $  \Omega_{K,0}/c_{s}^{2} $, where $  \Omega_{K,0} $ is today's fractional density of spatial curvature and $  c_{s} $ is the speed of sound during inflation. Here we study these effects in the limit $  c_{s}\ll 1 $. First, we show that the standard cosmological soft theorems in flat universes are violated in curved universes and the soft limits of correlators can have non-universal contributions even in single-clock inflation. This is a consequence of the fact that, in the presence of spatial curvature, there is a gap between the spectrum of residual diffeomorphisms and that of physical modes. Second, there are curvature corrections to primordial correlators, which are not scale invariant. We provide explicit formulae for these corrections to the power spectrum and the bispectrum to linear order in curvature in single-clock inflation. We show that the large-scale CMB anisotropies could provide interesting new constraints on these curvature effects, and therefore on $  \Omega_{K,0}/c_{s}^{2} $, but it is necessary to go beyond our linear-order treatment.}

\begin{document}
\maketitle
\flushbottom

\vspace{1cm}

\section{Introduction and summary}


The corroborated assumption that our universe is homogeneous and isotropic on large scales highly restricts the form of the spacetime metric. The only unknowns are the scale factor, whose evolution is dictated by Einstein equations, and the comoving spatial curvature $ K$, which is fixed by the boundary conditions. General relativity gives us no guidance in choosing a specific value of $ K$ and so it is important to derive predictions for observables for generic values of $ K$ and confront them with cosmological data. Current bounds constrain spatial curvature today to be at the per mille level or smaller \cite{Aghanim:2018eyx}
\begin{align}
\Omega_{K,0}\equiv \frac{K}{a_{0}^{2}H_{0}^{2}}=0.0007\pm 0.0019 \quad \text{(68$ \%$, Planck + BAO)}\,.
\end{align}
But our cosmological model gives us also an estimate for a \textit{lower bound} on the absolute value of $ \Omega_{K}$. This comes from the local effect of superHubble curvature perturbations. More specifically, it has been measured that subHubble curvature perturbations have an amplitude of about $ 2 \times 10^{-9}$ and an approximate scale-invariant spectrum. It is natural to expect that these perturbations continue to exist on superHubble scales. To leading order, the effect of these superHubble perturbations on our Hubble patch is to induce spatial curvature and a tidal force on Hubble scales (this follows from using Fermi normal coordinates \cite{Manasse:1963zz,Baldauf:2011} or more conveniently their cosmological generalisation known as conformal Fermi coordinates \cite{Pajer:2013ana,Dai:2015rda}). An estimate for the local spatial curvature in our Hubble patch due to superHubble fluctuations then is 
\begin{align}
|\Omega_{K}|\sim\ex{\Omega_{K}^{2}}^{1/2}&=  \frac{2}{3 H_{0}^{2}}\left[ \int_{0}^{H_{0}}\frac{d^{3}k}{(2\pi)^{3}}\frac{d^{3}k'}{(2\pi)^{3}}\ex{k^{2}\zeta(k)k'^{2}\zeta(k')} \right]^{1/2}\\
&=\frac{2}{3 H_{0}^{2}} \left[ \int_{0}^{H_{0}}\frac{ dk}{2\pi^{2}} k^{6} P(k) \right]^{1/2} \\
&= \frac{2}{3 H_{0}^{2}} \left[ \int_{0}^{H_{0}}dk  k^{3} \Delta_{\zeta}^{2}  \right]^{1/2} =\frac{\Delta_{\zeta}}{3}\simeq 1.5 \times 10^{-5}\,,
\end{align}
where we used the relation\footnote{This relation is valid only at linear order. But in standard cosmological models $  |\Omega_{K,0}|\ll 1 $ is actually an upper bound on the value of $ \Omega_{K}$ at any time during the hot big bang. We are therefore entitled to account for the effect of curvature during the hot big bang to linear order.} $  K=-(2/3)\nabla^{2}\zeta $. An anisotropic tidal field is expected at a comparable level. Even though the above explicit estimate assumes a scale invariant power spectrum on all superHubble scales, the integral is dominated by scales that are around the current Hubble radius and so is quite insensitive to changes in the spectral index for ultra-long wavelengths.

The effect of curvature on cosmological observables such as the Cosmic Microwave Background (CMB) or Large Scale Structures (LSS) has been well studied in the literature and many existing numerical Boltzmann codes already allow one to include spatial curvature in solving cosmological perturbation theory. All effects of curvature during the hot big bang are controlled by the parameter $ \Omega_{K}(t)$, evaluated at the relevant time for the given observable. As is well-known, $ \Omega_{K}(t) $ is an increasing function of time in decelerated cosmologies. As a consequence, what controls the effect of curvature in the late universe is bounded by the value of curvature today $ \Omega_{K}(t) \lesssim \Omega_{K,0}$.

\begin{figure}
\centering
\includegraphics[width=\textwidth]{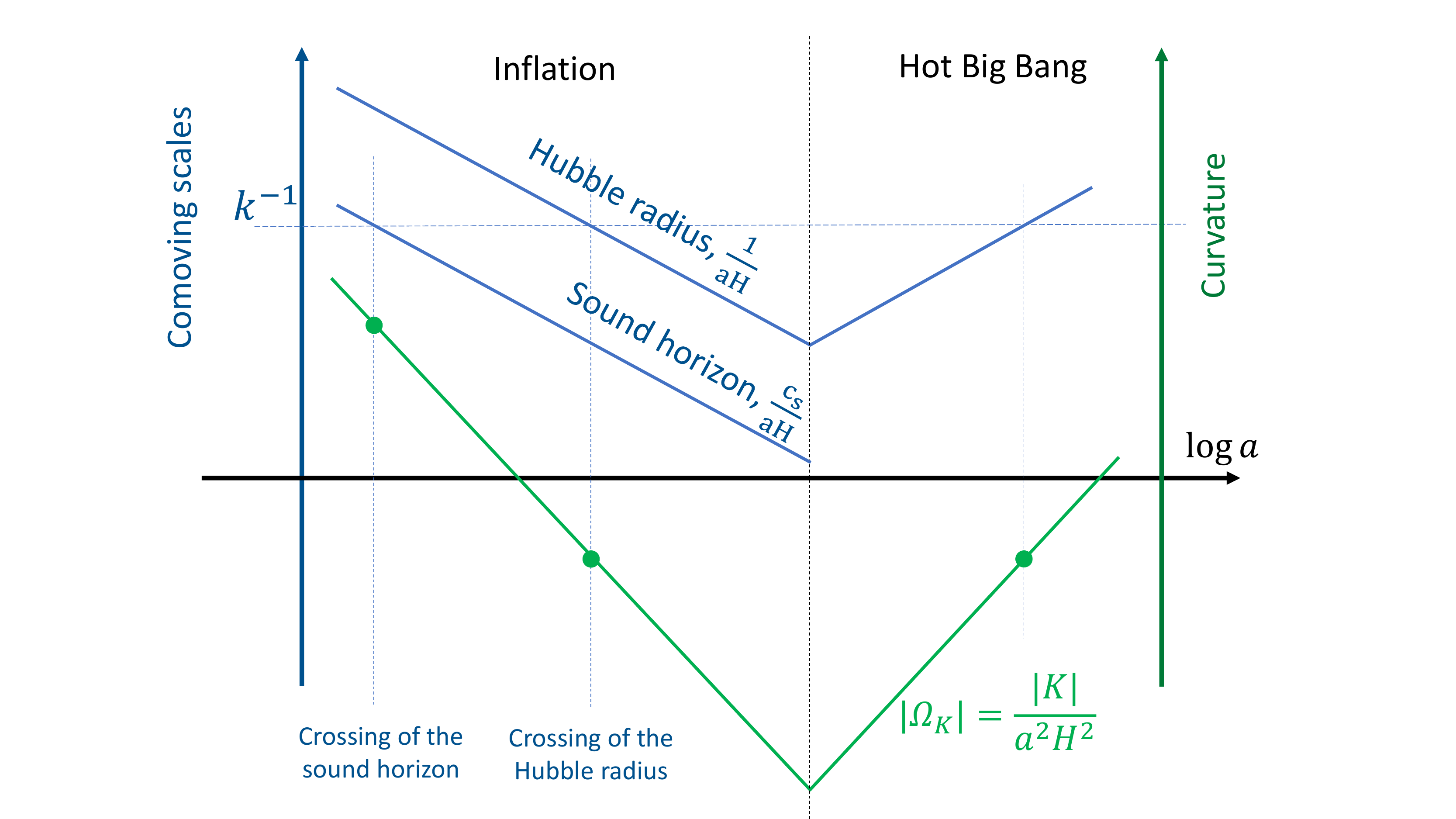}
\caption{The figure shows why primordial perturbations are sensitive to $  \Omega_{K,0}/c_{s}^{2} $. Since the fractional density of spatial curvature (green line) decreases during inflation, it is larger at sound-horizon crossing (first green dot) than at the crossing of the Hubble radius (second green dot). \label{fig1}}
\end{figure}

What motivated this paper is the observation that the effect of curvature on primordial perturbations from inflation is controlled instead by $ \Omega_{K,0}/c_{s}^{2}$. Therefore, if primordial perturbations had a small speed of sound, $  c_{s}\ll 1 $, then they could provide a very sensitive probe of spatial curvature. This dependence on $ c_{s}$ is easy to understand (see Fig. \ref{fig1}). During inflation, perturbations freeze out at the so-called sound horizon, i.e. when the comoving wavenumber satisfies $ c_{s} k = a H$. Consider now a perturbation of size the Hubble radius today $ k_{H_{0}} = a_{0} H_{0} = H_{0} $. The value of $ \Omega_{K}$ at the time during inflation when $ k_{H_{0}} $ froze out is 
\begin{align}\label{cs}
\Omega_{K}\Big|_{\text{freeze}}=\frac{K}{a^{2}H^{2}}\Big|_{\text{freeze}}=\frac{K}{c_{s}^{2}k^{2}_{H_{0}}}=\frac{\Omega_{K,0}}{c_{s}^{2}}\,.
\end{align}
So, if $ c_{s}\ll 1$ primordial perturbations felt a much larger value of $ \Omega_{K}$ right before they stopped evolving than any cosmological observable in the late universe. The current bound\footnote{A more detail discussion of the bound on $  c_{s} $ and the other EFT free parameter at the same order will be given in Sec. \ref{5p1}.} on $ c_{s}$ is 
\begin{align}
c_{s}\geq 0.021 \quad \text{(95$\% $, Planck T+E)}\,.
\end{align}
Therefore we expect that the factor $ c_{s}^{-2}$ in \eqref{cs} will give us a boost of up to a thousand in sensitivity to spatial curvature. The explicit calculations in this paper confirm this rough estimate and furthermore show that the final observable effect also depends on the strength of interactions during inflation. For a fair comparison with other probes of curvature, it should also be mentioned that the effects of curvature on primordial correlators peak at the largest observable scales, where cosmic variance is largest. While this heuristic argument applies also to multifield inflation, in this work we focus exclusively on single-field inflation.

In this paper, we take the observation that a small speed of sound enhances the sensitivity to spatial curvature in two distinct but related directions. First we study how spatial curvature can affect the soft theorems for cosmological correlators, which in a flat universe provide model-independent consistency relations to test the assumption of a single clock during inflation. While theoretical predictions for correlators are highly model dependent, in recent years it has become clear that symmetries, shared by large classes of models, lead to specific predictions known as soft theorems, which can be tested with current and upcoming data. Soft theorems constrain the squeezed limit of correlators, in which one of the momenta of the correlator is much smaller than any relevant scale in the problem. Cosmological soft theorems take the schematic form
\begin{align}\label{softth}
\lim_{\v{q}\to 0}\frac{\ex{ \O (\v{q})\O (\v{k}_{1})\dots \O (\v{k}_{n}) }'}{\ex{\O (\v{q})\O (\v{q}) }'} = \sum_{a=1}^{n} L_{a} \ex{\O  (\v{k}_{1}) \dots \O (\v{k}_{n}) }'\,,
\end{align}
where $  \O $ are some operators, a prime denotes that we have dropped the momentum-conserving delta function and $  L=L(k,\partial_{k}) $ is some linear operator consisting of powers of the momenta and derivatives. The most famous soft theorem has been derived by Maldacena in \cite{Maldacena:2002vr} and fixes the squeezed bispectrum in terms of the power spectrum and it applies to all attractor, single-field models of inflation \cite{Creminelli:2004yq}. This first result has been extended to higher n-point functions for primordial scalar, tensor and vector perturbations \cite{Hinterbichler:2012nm,Hinterbichler:2013dpa,Assassi:2012zq,Creminelli:2011rh,Creminelli:2012ed,Creminelli:2013cga,Pimentel:2013gza,Berezhiani:2013ewa,Berezhiani:2014tda,Bordin:2016ruc,Pajer:2017hmb,Hui:2018cag}. Soft theorems are conveniently interpreted as the consequence of residual, non-linearly realized symmetries associated with adiabatic modes \cite{Weinberg:2003sw,Hinterbichler:2013dpa,Mirbabayi:2014hda,Pajer:2017hmb,Finelli:2017fml}, namely physical perturbations that are indistinguishable from a change of coordinates in the neighborhood of a point in spacetime. In the presence of additional symmetries beyond diffeomorphism invariance, new adiabatic modes and new soft theorems can be derived. One example are non-attractor models of inflation such as Ultra-Slow-Roll inflation \cite{Kinney:2005vj} that are also invariant under a shift symmetry. In this setup, new \textit{generalized adiabatic modes} can be found, which are locally indistinguishable from a change of coordinates \textit{and} a symmetry transformation \cite{Finelli:2017fml,Finelli:2018upr}. The violation of Maldacena's consistency relation in shift-symmetric Ultra-Slow-Roll inflation \cite{Namjoo:2012aa,Chen:2013aj,Akhshik:2015rwa,Mooij:2015yka}, can be attributed to the fact that cosmological perturbations asymptote generalized adiabatic modes, as opposed to the standard adiabatic modes \cite{Finelli:2017fml,Bravo:2017wyw}. Adiabatic modes and their associated soft theorems can also be derived \cite{Bordin:2017ozj,Pajer:2019jhb} in the presence of alternative spacetime symmetry breaking patterns as in solid inflation \cite{Gruzinov:2004ty,Endlich:2012pz} and other generalizations \cite{Bartolo:2015qvr,Ricciardone:2016lym,Piazza:2017bsd,Fujita:2018ehq}. All results so far have been obtained in spatially-flat FLRW spacetimes\footnote{The only exception known to us is \cite{Xiao:2014zna}, where the author studies soft theorems in a ``toy'' closed universe in 2+1 dimensions. The overall scaling of the violation of the consistency relation observed in that work in around Eq. (20) seems to match our results in 3+1 dimensions.}. In this work, we study adiabatic modes and soft theorems in spatially-curved universes. A summary of our result can be found in the next subsection.

A second direction in which we push our investigation is the explicit calculation of corrections to the power spectrum and bispectrum of curvature perturbations that are induced by spatial curvature at linear order in $ \Omega_{K}$. To achieve this result we take advantage of the fact that, at linear order, the local effect of spatial curvature is the same as that of a suitable isotropic long wavelength perturbation on a spatially flat background. This equivalence allows us to use results in the literature for the bispectrum and trispectrum in flat FLRW spacetime to deduce the linear order effect of curvature on the power spectrum and bispectrum. The calculation simply amounts to extract a specific term from the squeezed limit of the higher-point correlator. Our main findings are summarized in the next subsection.

The rest of the paper is organized as follows. In the next subsection we give a brief summary of our main results. In Sec. \ref{sec:AM} we outline the procedure to derive adiabatic modes in curved FLRW universes and discuss the difficulty in continuing these modes to physical momentum. In Sec. \ref{sec:SoftThms}, after reviewing the derivation of soft theorems in flat universes, we show that the standard ``monochromatic'' soft theorems do not exist in the presence of spatial curvature. This holds both for soft scalar and soft tensor modes. Here we also briefly mention some other non-standard and less phenomenologically relevant ways to constrain the soft behavior of correlators. In Sec. \ref{sec:SqueezedLimits} we calculate the theoretical prediction for the effect of curvature on the power spectrum and bispectrum of curvature perturbations, both for canonical single-field inflation and for the decoupling limit of the Effective Field Theory of Inflation \cite{Cheung:2007st}. In Sec. \ref{sec:ObsSignatures}, we show that, giving current constraints, the correction to the power spectrum can be large enough to be measurable in the CMB, but this happens in a regime in which one needs to go beyond our linear treatment of curvature. We also make some estimates of the how large the non-scale invariant corrections to the bispectrum could be, given current constraints. Finally we conclude in Sec. \ref{sec:Discussion} with a discussion and an outlook.

\textbf{Notation and conventions}: We use a mostly positive signature. Greek indices from the middle of the alphabet run over $  \mu,\nu=0,1,2,3 $ and latin indices from the middle of the alphabet over $ i,j=1,2,3$. We define symmetrization of indices by $a_{(ij)}=\frac{1}{2}(a_{ij}+a_{ji})$. Our Fourier conventions are
\begin{align}
f(x)=\int \frac{d^{3}k}{(2\pi)^{3}}e^{i \v{k} \v{x}} f(\v{k})\,, \quad\quad f(k)=\int d^{3}x \,e^{-i \v{k} \v{x}} f(\v{x})\,.
\end{align}
Spatial 3-vectors are indicated in boldface, as for example in ``$  \v{x} $'', and a hat denotes a unit norm vector $  \hat{\v{q}}\cdot \hat{\v{q}}=1 $.

\subsection{Summary of the results}

In the following we give a short summary of our main results for soft theorems and for the curvature corrections to the power spectrum and bispectrum.

\paragraph{Soft theorems} We find that the \textit{standard soft theorems that relate $  n $-point to $  (n-1) $-point functions in the squeezed limit are generally violated in curved universes}. In particular the correlation in the squeezed limit is not simply a change of coordinates and therefore does not have the same universal character in a curved universe that it has in a flat universe. In attempting to reproduce the flat universe derivation of soft theorems, we derived all residual diffeomorphisms (diffs) in Newtonian gauge. Residual diffs, for both scalar and tensor modes do exist in curved universes and reduce to the respective flat-universe adiabatic modes in the $  K\to 0 $ limit. The main obstacle emerges when one tries to connect residual diffs to physical modes. For both scalars and tensors and in both open and closed universes the spectrum of physical modes (i.e. the eigenvalues of the Laplacian) is separated from the spectrum of residual diffs by a discrete gap. As a consequence of this, the time evolution of physical modes is different from that of residual diffs already at linear order in curvature. When deriving soft theorems, one substitutes physical modes with diffs in some soft limit of a correlator. This introduces an error already at linear order in curvature and so we conclude that soft theorems are violated by curvature corrections. The violation is parameterized by $  \Omega_{K,0}/c_{s}^{2} $ but depends also on the strength of the interaction of perturbations, which are dictated by an explicit inflationary model or parameterized by the EFT of inflation. For the EFT of Inflation our main result for the squeezed bispectrum of curvature perturbations is the expression in \eqref{final2}, while for canonical single-field inflation we find \eqref{eq:BispectrumCanonicalInflation}.

\paragraph{Power spectrum and bispectrum} To linear order, the local effect of spatial curvature can be traded for that of a long wavelength curvature perturbation (see e.g. \cite{Baldauf:2011,Creminelli:2013cga,Dai:2015rda}). This fact was used in \cite{Creminelli:2013cga} to show that the terms at order $  k_{L}^{2} $ in the squeezed bispectrum (i.e. for $  k_{L}\to 0 $) are related to the corrections of spatial curvature to the power spectrum. These corrections are not scale invariant and peak on the largest scales, see \eqref{eq:PScurved} and \eqref{uno}-\eqref{pidot}. We confront the curvature-dependent power spectrum prediction with data on the CMB temperature anisotropies. The signal-to-noise is saturated by just the first few $  C^{TT}_{l} $'s. The bounds that we derive (see Figure \ref{fig:powervsPlanck}) are slightly weaker than the theoretical bounds we have from the validity of our linear treatment of curvature. This means that the CMB can potentially improve current bounds on the $  \{\Omega_{K,0},c_{s}\} $ plane in the direction $  \Omega_{K,0}/c_{s}^{2} $, but this requires computing the correction to the power spectrum to all orders in $  K $. We also compute curvature corrections to the bispectrum from the squeezed trispectrum in canonical single-field inflation (from \cite{Seery:2006,Seery:2008ax}), finding \eqref{eq:BispectrumCanonicalInflation}, and for the so-called $  P(X) $-theories \cite{Chen:2009se} (equivalent to the leading terms in the EFT of inflation), finding \eqref{uno}-\eqref{tre}. We show that the the signal-to-noise for this leading order curvature correction to the bispectrum is at most of order one for the allowed values of parameters and always smaller than one within the regime of validity of our analysis.

 
\section{Residual diffeomorphism}\label{sec:AM}

In this section, we derive residual diffs in spatially-curved FLRW universes. Our main finding is that residual diffs do exist, but their momenta are always separated from the spectrum of physical modes by a discrete amount.



\subsection{Gauge fixing}\label{ssec:AM:gauge_fixing}

In this work, we consider curved FLRW universes with the following spacetime metric 
\begin{equation} \label{flrwm}
ds^2 = -dt^2 + a^2(t) \tilde g _{ij}(\bfx) dx^i dx^j \equiv \bar g_{\mu\nu}dx^\mu dx^\nu,
\end{equation}
where 
\begin{equation}
\tilde g _{ij} (\bfx) = f^2(K\bfx^2) \delta_{ij}\,\qquad \text{with}\qquad f(K\bfx^2)\equiv \frac 1 { \left( 1 + \frac 1 4 K \bfx^2 \right)}\,,
\end{equation}
and we have defined $\bfx^2\equiv\delta_{ij}\,x^ix^j$.
The universe is spatially flat, open or closed if $K=0,K<0$ and $K>0$, respectively. In the open case, the radial coordinate satisfies $\bfx^2 < 4/|K|$, while in the closed case $ 0\leq x \leq + \infty$. The volume contained in an open universe is infinite, while it is finite in a closed universe.
 Other useful properties of the FLRW metric can be found in Appendix \ref{app:FLRW}.

For the perturbations around the FLRW background, 
we choose to use the Newtonian gauge, which is defined through
\begin{equation}
\label{NG}
ds^2 = -(1+2 \Phi)dt^2 + a G_i dt dx^i + a^2 \left[ (1-2\Psi) \tilde g _{ij} + \gamma_{ij} \right] dx^i dx^j\,,
\end{equation}
with
\be
\label{trtl}
\nabla_i\,G^i=\nabla_{i}\,\gamma^{ij}=\gamma^i_{\,i}=0\,.
\ee
Here, $\nabla_i$ is the covariant derivative with respect to the spatial metric $\tilde{g}_{ij}$. We raise and lower spatial indices with the $\tilde{g}_{ij}$ metric. Metric perturbations are denoted by $h_{\mu\nu}$. We take the energy momentum tensor to be that of a single perfect fluid, so a universe with a single scalar field is also included in our study. To first order in perturbations, this implies
\be
T^{\mu \nu} &=& (\rho + p) u^\mu u^\nu + g^{\mu \nu} p\,, \\ \nn
\rho&=&\bar{\rho}(t)+\delta\rho\,,\\ \nn
p&=&\bar{p}(t)+\delta p\,,\\ \nn
u_\mu  &=&(-1+\dfrac{1}{2}h_{00},\nabla_i\,\delta u+u_i^V), \quad  \tilde{g}^{ij}\nabla_i\,u_j^V  =  0 .
\ee
Generally, there might be residual diffeomorphisms that are compatible with the gauge choice.\footnote{It should be possible to avoid such residual diffeomorphisms by an appropriate choice of coordinates. In flat space, even in such coordinates, cancelations among scalar, vector and tensor perturbations in the zero momentum limit lead to a set of adiabatic modes. See \cite{Pajer:2019jhb} for a related discussion in the context of Solid Cosmologies.} Under an infinitesimal change of coordinates $x^\mu \to x^\mu + \epsilon^\mu$, metric perturbations transform as
\begin{align}
\Delta h_{00}  =& 2 \dot{\eps}^0\,, \\ 
\Delta h_{0i} =& \partial_i \eps^0 - a^2 f^2 \dot \eps^i  ,\\
\Delta h_{ij} =&- 2 H \eps^0 \gb_{ij} + K  f x^k \eps^k \gb_{ij} - 2\gb_{k(i} \partial_{j)} \eps^k\,,
\end{align}
which in turn yield
\be
\label{perts}
&& \Phi=-\dot{\eps}^0\,,\\ \nn
&&\Psi=H\eps^0-\dfrac{1}{2}K\,f\,x^k\eps^k+\dfrac{1}{3}\partial_k\eps^k\,,\\ \nn
&& G_i =\nabla_i\eps^0-a^2\,\dot{\eps}_i\,,\\ \nn
&&\gamma_{ij}=2\nabla_{(i}\eps_{j)}-\dfrac{1}{3}\nabla_k\eps^k\,\tilde{g}_{ij}\,.
\ee
The variables parameterizing the perfect fluid on the other hand change as 
\begin{align}
\label{matt}
\Delta \delta \rho =&-\dot{\bar{\rho}}\epsilon^0,\\ \nn
\Delta \delta p =&-\dot{\bar{p}}\epsilon^0,\\ \nn
\Delta \partial_i\,\delta u+\Delta u^V_i =&\partial_i \epsilon^0\,.
\end{align}
To maintain the Newtonian gauge, we must impose \eqref{trtl}, giving
\be
\label{resdifs}
&& \nabla_i\,G^i=0 \Rightarrow \quad \nabla^2\eps^0-a^2\nabla_i\dot{\eps}^i =0\,,\\
\label{tensdis}
&&\nabla^i\gamma_{ij}=0\Rightarrow\,\quad \nabla^i\Big(2\nabla_{(i}\eps_{j)}-\dfrac{1}{2}\nabla_k\eps^k\,\tilde{g}_{ij}\Big)=0\,.
\ee

For diffs that respect \eqref{resdifs} and \eqref{tensdis}, general covariance guarantees that the perturbations in \eqref{perts} solve the linearized Einstein equations. However, just as for flat FLRW, for these diffs to have a chance to approximate physical perturbations, additional conditions must be satisfied. To see this, recall that physical perturbations in curved universes can be uniquely decomposed into scalars, vectors and tensors (with appropriate fall-off conditions in the open case, see Appendix \ref{app:FLRW}). Then let us decompose the linearized Einstein equations, $ \delta_{\mu\nu}=0$, in the following way,
\begin{eqnarray}
&&\delta E_{00}=S^{(1)}\,,\\ \nn
&&\delta E_{0i}=\nabla_i\,S^{(2)}+V^{(1)}_i\,,\\ \nn
&&\delta E_{ij}=S^{(3)}\bar{g}_{ij}+\nabla_i\nabla_j\,S^{(4)}+\nabla_{(i}V^{(2)}_{j)}+T_{ij}\,,
\end{eqnarray}
where, $S^{(i)}$, $V^{(i)}$ and $ T_{ij}  $ are the scalar, transverse vector and transverse traceless tensor components, respectively. For physical perturbations $  S^{(1,2,3,4)}=V^{(1,2)}_{i}=T_{ij}=0 $. These equations are then \textit{necessary} conditions (but as we will see not sufficient) for any residual diffs to be able to approximate physical perturbations. Since for  residual diffs we already know that $\delta E_{\mu\nu}=0 $, we need only to further impose\footnote{Notice that after setting $  S^{(4)}=0 $, Einstein's equations automatically imply $  S^{(3)}=S^{(1)}=0 $.}
\be
\label{adc}
S^{(2)}=S^{(4)}=V^{(2)}=0\,,
\ee
where when we had the choice we set to zero those components with at most one time derivative. It is easy to verify that the rest of the components, namely $S^{(1)}$, $V^{(1)}$, $S^{(3)}$, and $T$, must also vanish as result of general covariance and rotational symmetry. In Newtonian gauge, \eqref{adc} becomes
\be
\label{adiabaticity}
&& S^{(2)}:\qquad
\dot{\Psi}+H\Phi=\left( \dot{H}-\dfrac{K}{a^2}  \right)\delta u\,,\\ 
&& S^{(4)}:\qquad \Phi =\Psi\,.\label{adiab2}
\ee
We will refer to these equations as ``adiabaticity conditions''. Notice that there are no adiabaticity conditions for tensor residual diffs. Since vector modes decay in standard cosmologies, we set them to zero (so that $V^{(2)}=0$) and ignore them in the rest of the paper.


\subsection{Scalar residual diffs}\label{ssec:AM:scalar}

In this subsection, we investigate the existence of scalar residual diffs in an open or closed universe that satisfy the adiabaticity conditions \eqref{adiabaticity} and \eqref{adiab2}. Demanding $\gamma_{ij}$ to vanish restricts $\epsilon^i$ to solutions of the following equation
\be
\D_i\eps^j+\D_j\eps^j=\dfrac{2}{3}\delta_{ij}\,\D_k\eps^k\,. 
\ee
This is nothing but the conformal Killing equation in Euclidean space. It has the following solutions
\be
&(\text{Dilation})&\quad \eps^i_{\text{d}}=\lambda(t)x^i\,, \label{eq:DilDiff} \\ \label{scteq}
&(\text{Special conformal transformation})&\quad  \eps^i_{\text{SCT}}=\textbf{b}(t).\bfx\, x^i-\dfrac{1}{2}\bfx^2\,b^i(t)\,,\\ 
&(\text{Translation})&\quad \eps^i_{\text{T}}=c^i(t)\,,\\ 
&(\text{Rotation})&\quad \eps^i_{\text{R}}=\omega_{ij}(t)\,x^j\,.
\ee
This is a good point to pause and discuss the spatial profiles of these residual diffs. Translations and rotations are isometries of the background metric and so do not affect perturbations. Dilations and special conformal transformations on the other hand do change perturbations. A common feature that is true regardless of spatial curvature is that the residual diffs in \eqref{eq:DilDiff} and \eqref{scteq} cannot be fixed by imposing local conditions on the matter fields and metric components. Let discuss some additional curvature-dependent properties:
\begin{itemize}
\item In flat space, these residual diffs do not fall off at spatial infinity ($  |\v{x}|\to +\infty $) and are therefore known as ``large'' diffs. This behavior should be contrasted with that of physical perturbations that are required to vanish at spatial infinity (so as to justify neglecting total spatial derivatives).
\item In open universes, spatial infinity coincides with $|\bfx|\to 2/\sqrt{|K|}$, and residual diffs do \textit{not} vanish there. In this sense these diffs could also be called ``large'' diffs. Again this should be contrasted with physical perturbations that should vanish as $|\bfx|\to 2/\sqrt{|K|}$. 
\item In closed universes, the spatial maximally-symmetric manifold is compact and there is no spatial infinity. In this case the above residual diffs are regular everywhere and they are square integrable. In this sense they are not ``large" gauge transformations. They are only residual in the sense that they cannot be fixed by local gauge conditions.
\end{itemize}  

In this work we focus on finding a counterpart to Weinberg's first adiabatic mode \cite{Weinberg:2003sw}, which in flat space can be generated by a time-independent dilation, $\eps^i_{\text{d}}=\lambda\,x^i$. Unlike for the flat case, in curved universes the coefficient $\lambda$ might have nontrivial time dependence, which is fixed by imposing the adiabaticity conditions in \eqref{adiabaticity} and \eqref{adiab2}.

The absence of vector modes means that $G_i$ in \eqref{perts} must vanish. This implies
\begin{equation}
\eps^0 = - \frac{2a^2 }{K} f \dot{\lambda} + \mathcal D (t) ,
\label{eq:TimeDiff}
\end{equation}
where $\mathcal D$ is an integration ``constant''. Inserting $\eps^0$ and $\eps^i$ into \eqref{matt} and \eqref{perts}, and assuming no shift symmetry on $\delta u$\footnote{For a perfect fluid, $\delta u$ always has a shift symmetry. This simply reflects the equivalence between perfect fluids and superfluids ($P(X)$ theories) in the limit of no-vorticity. We are however interested in a generic single scalar field cosmology, where the relation between $\delta u$ and the scalar perturbation $\delta \phi$ breaks the shift symmetry \cite{Pajer:2017hmb}. 
}
, we find the following solution
\be
\label{pscng}
\Phi&=&\dfrac{1}{K}\dfrac{2\partial_t(a^2\dot{\lambda})}{1+\frac{1}{4}K\bfx^2}-\dot{{\cal D}}(t)\,,\\ \nn
\Psi&=&H{\cal D}-\lambda+\dfrac{2(\lambda-\frac{H}{K}a^2\dot{\lambda})}{1+\frac{1}{4}K\bfx^2}\,,\\ \nn
\delta u&=& - \frac{2a^2 }{K} f \dot{\lambda} + \mathcal D (t)\,.
\ee
So far, $\lambda$ and ${\cal D}$ have been arbitrary time-dependent functions. They are fixed however by imposing the adiabaticity conditions \eqref{adiabaticity} and \eqref{adiab2}:
\be
&&{\cal D}=\dfrac{a^2\dot{\lambda}}{K}\,,\\ 
&&\ddot{\lambda}+3H\dot{\lambda}-\dfrac{K}{a^2}\lambda=0\,.
\label{eq:LambdaTimeDep}
\ee
In summary, we have found the following seemingly adiabatic solution
\be
\label{wamI}
\Phi&=&\Psi=\dfrac{1}{K}\partial_t (a^2\dot{\lambda})\left(\dfrac{2}{1+K\bfx^2/4}-1\right)\,,\\ \nn
\dfrac{\delta \rho}{\dot{\rho}}&=&\dfrac{\delta p}{\dot{p}}=-\delta u=\dfrac{a^2\dot{\lambda}}{K}\left(\dfrac{2}{1+K\bfx^2/4}-1\right)\,, \\ \nn
{\cal R}\equiv-\Psi+H\delta u&=&-\lambda\,\left(\dfrac{2}{1+K\bfx^2/4}-1\right)=-\Psi-H\frac{\delta \rho}{\dot{\bar \rho}}\equiv \zeta\,.
\ee
In the last line, ${\cal R}$ and $  \zeta $ are the curvature perturbations on comoving and constant-density slices, respectively. In the flat-space limit, $K \to 0$, both Weinberg's first and second adiabatic modes are obtained from the two independent $\lambda(t)$ solutions to \eqref{eq:LambdaTimeDep}.

For future reference, notice that 
\begin{align}
\nabla^2 \zeta = f^{-2} \left( \partial_i \partial_i - \frac{1}{2} K f x^k \partial_k \right) \zeta= - 3 K \zeta\,.
\end{align}  
That is, our scalar residual diff has $\nabla^2 = - 3K$ (see Table \ref{tab1}).

As a consistency check, we note that the evolution equation for scalar perturbations $\mathcal{R}$ in curved universe with a scalar inflaton field, as derived for example in \cite{Handley:2019}, takes the form
\begin{align} \nn
(\mathcal{D}^2 - K \mathcal{E}) \ddot{\mathcal{R}} & + \left[ \left( H + 2 \frac{\dot{z}}{z} \right) \mathcal{D}^2 - 3KH \mathcal{E} \right] \dot{\mathcal{R}} \\
& + \frac{1}{a^2} \left[ K \left(   1 + \mathcal{E} -\frac{2}{H} \frac{\dot{z}}{z} \right) \mathcal{D}^2 - \mathcal{D}^4 + K^2 \mathcal{E} \right] \mathcal{R} = 0,
\label{eq:ScalarTimeDep}
\end{align}
where
\be
 \mathcal{D}^2 \equiv \nabla^2 + 3K,  \quad  z = \frac{a \dot{\phi}}{H},  \quad	\mathcal{E} = \frac{\dot{\phi}^2}{2 H^2}.
\ee
The limit $\mathcal{D}^2 \to 0$, which corresponds to the residual diffs, gives
\be
\ddot{\mathcal{R}} + 3H  \dot{\mathcal{R}} - \frac{K}{a^2} \mathcal{R} = 0,
\ee
confirming (\ref{eq:LambdaTimeDep}).


\subsection{Tensor residual diffs}\label{ssec:AM:tensors}

We switch now to tensor perturbations for which there is no adiabaticity condition. To remove scalar modes we set $\delta u=\Phi=\Psi=0$. This results in $\epsilon^0=0$ and
\be
\label{trep}
\nabla_i\epsilon^i=0\,.
\ee
In addition, setting $G_i=0$ enforces $\epsilon^i$ and, subsequently, $\gamma_{ij}$ to be time-independent. 
The spatial diffs must further satisfy the gauge condition \eqref{tensdis}, 
\be
\label{tediff}
\nabla^i\,\Big(\nabla_i\eps_j+\nabla_j\eps_i\Big)=0\,.
\ee
It is straightforward to see that
\be
\label{lapeps}
\nabla^2\,\eps^i =-2K\,\eps^i \then \label{lapten}
 \nabla^2\,\gamma_{ij}=+2K\gamma_{ij} \quad \text{(tensor residual diffs)}\,.
\ee
Notice that \eqref{lapten} is compatible with Einstein's equations for physical modes, which for tensors lead to
\be
\ddot{\gamma}_{ij}+3H\dot{\gamma}_{ij}-\dfrac{1}{a^2}(\nabla^2-2K)\gamma_{ij}=0 \quad \text{(physical modes)}\,.
\ee
From this we see that a tensor residual diff, which must be constant in time, indeed satisfies 
\begin{align}
\ddot \gamma_{ij}=\dot \gamma_{ij}=0 \quad \Rightarrow \quad \nabla^2 \gamma_{ij}=+2K \gamma_{ij}\,. 
\end{align}
Just like their analogues in flat space, equations \eqref{lapeps} and \eqref{trep} admit infinitely many solutions \cite{Tanaka:1997kq}. These are easier to be written in spherical coordinates, where the solutions to \eqref{lapeps} are either of even (+) or odd (-) parity. They are found to be (see \cite{Tanaka:1997kq})
\begin{align}
\bar{\eps}_r^{(+)lm}&=V_1^l(r)\,Y_{lm}(\theta,\phi)\,,& \bar{\eps}^{(+)lm}_\theta&=V_2^l(r)\partial_{\theta}Y_{lm}(\theta,\phi)\,,& \bar{\eps}^{(+)lm}_\phi&=V_2^l(r)\partial_{\phi}Y_{lm}(\theta,\phi)\,, \\ \nn
\bar{\eps}_r^{(-)lm}&=0\,,& \bar{\eps}^{(-)lm}_\theta&=V_3^l(r)\partial_{\theta}Y_{lm}(\theta,\phi)\,,& \bar{\eps}^{(-)lm}_\phi&=V_3^l(r)\partial_{\phi}Y_{lm}(\theta,\phi)\,.
\end{align}
Above, we have pedantically denoted the spatial diffs in the spherical coordinates with $\bar{\eps}_a\,(a=r,\theta,\phi)$ to distinguish them from their Cartesian counterparts, and we have defined 
\be
V_1^l(r)&=&\dfrac{1}{\sqrt{r^3\,f(Kr^2)}}\,P^{-l-1/2}_{-5/2}\Big(\dfrac{4-Kr^2}{4+Kr^2}\Big)\,,\\ \nn
V_2^l(r)&=&\dfrac{1}{l(l+1)}\dfrac{1}{f(Kr^2)}\partial_r \left[ \sqrt{r\,f(Kr^2)}P^{-l-1/2}_{-5/2}\left( \dfrac{4-Kr^2}{4+Kr^2} \right) \right]\,,\\ \nn
V_3^l(r)&=&\sqrt{r\,f(Kr^2)}P^{-l-1/2}_{-5/2}\left( \dfrac{4-Kr^2}{4+Kr^2} \right)\,,
\ee
where $P^\mu_\lambda$ are associated Legendre functions. The resulting tensor modes can be computed by inserting any of the above $\bar{\epsilon}_a$'s in 
$\bar{\gamma}_{ab}=2\nabla_{(a}\bar{\epsilon}_{b)}$.

 
\subsection{From large diffs to physical modes}\label{ssec:}

Now we would like to see if the residual diffs satisfying the adiabaticity conditions that we have found in the previous sections can be smoothly connected to physical modes. In a flat universe this is the case, as in the $ k\to 0$ limit physical perturbations become indistinguishable from (large) residual diffs, for which $ k=0$. As we will see below, in spatially curved universes we find an obstruction in the form of a discrete gap between the wavenumber of residual diffs and the spectrum of physical modes. This is summarised in Table \ref{tab1}. 

\begin{table}[t]
\centering
{\setlength{\extrarowheight}{5pt}
\begin{tabular}{|c|c|c|c|}
\hline
\multicolumn{2}{|c|}{}                 & Open ($K < 0$)                  & Closed ($K > 0$)                           \\ \hline
\multirow{2}{*}{Scalars} & Res. diffs & $3|K| > 0$                      & $-3K < 0$                                  \\ \cline{2-4} 
                         & Physical    & $-(1+p^2) |K| \leqslant - |K|$  & $-p(p+2) K < 0, \quad p = 2, 3, \ldots$    \\ \hline
\multirow{2}{*}{Tensors} & Res. diffs & $-2 |K| < 0$                        & $2 K > 0$                                  \\ \cline{2-4} 
                         & Physical    & $-(3+p^2) |K| \leqslant -3 |K|$ & $- (p(p+2) - 2) K < 0, \quad p = 2, 3, \ldots$ \\ \hline
\end{tabular} }
\caption{This table summarises the eigenvalues of the 3D spatial Laplacian $  \nabla^{2} $ for residual diffs (``res. diffs'') and physical modes (``physical'') in open and closed universes.\label{tab1}}
\end{table}

In both open and closed universes, the scalar and tensor residual diffs discussed in Sec. \ref{ssec:AM:scalar} and Sec. \ref{ssec:AM:tensors} are eigenfunctions of the $\nabla^2$ operator, with eigenvalues
 \begin{align}
 \nabla^{2}S(x)&= - 3 K S(x) & \text{(Scalar residual diffs)}\, , \label{LDSL}\\
  \nabla^{2}T(x)&= + 2 K T(x) & \text{(Tensor residual diffs)}\, .\label{LDTL}
 \end{align}
Let us compare this with the physical spectrum in open and closed universes. In an open universe ($  K<0 $), monochromatic\footnote{By a monochromatic mode, we mean any mode which is an eigenfunction of $\nabla^2$. Gradients $  \nabla_{i} $ do not commute with each other and therefore cannot be simultaneously diagonalized.} perturbations that provide a complete basis of square integrable functions consist of the so-called \textit{subcurvature} modes, all of which have negative eigenvalues  \cite{Lyth:1995cw, Garriga:1998he}:
 \begin{align}
 \nabla^{2}S_{plm}(x)&= -(1+p^{2})|K| S_{plm}(x),  & \text{(physical Scalars for $  K<0 $)}\,,\\
  \nabla^{2}T_{plm, ij}(x)&= - (3 + p^2) |K| T_{plm, ij}(x),  & \text{(physical Tensors for $  K<0 $)}\,,
 \end{align}
where $  p>0 $. Due to the existing gap between the momenta of the physical perturbations and the residual diffs, namely 
\begin{align}
-(1+p^{2})|K|<-|K|&\text{ vs } +3 |K| & &\text{(scalar gap)}\,,\\
- (3 + p^2)|K|<-3|K|&\text{ vs } -2|K|  & &\text{(tensor gap)}\,,
\end{align}  
monochromatic physical modes cannot capture the time dependence of the gauge modes in any continuous limit - neither for the scalar nor for the tensor. (This is in contrast with flat space, where eigenfunctions of $\nabla^2$ can have the eigenvalue $-k^2$ arbitrarily close to zero and asymptote to the behavior of the (large) residual diff in the long wavelength limit.)

In closed universes, $  K>0 $, residual diffs have Laplacian eigenvalues again given by \eqref{LDSL} and \eqref{LDTL}. Normalizable modes on the other hand obey
\begin{align}
\nabla^2 S_{plm}(x)&=- p(p+2)K S_{plm}(x) & \text{with }p&=0,1, \ldots ,\\
\nabla^2 T_{plm, ij} (x) &= -(p(p+2) - 2) K T_{plm, ij} (x)  & \text{with } p&=0, 1, \ldots.
\end{align}
However, all modes with $p = 0, 1$ are equivalent to diffs and physical modes only start at $p = 2$, so the modes we found still cannot be approached in any continuous way by monochromatic modes (see Sec. \ref{sec:SoftThms} for more discussion). 

The discussion so far has focused on monochromatic modes and as such we cannot preclude the possibility of reaching an adiabatic mode as a limit of some other non-monochromatic physical perturbation. We would like to find those perturbations that have the following property:
\begin{center} 
A family of physical perturbations parameterized by $\alpha$ is \textit{adiabatic} in the limit $\alpha \to 0$ if and only if they converge pointwise in spacetime to a residual diff as $\alpha \to 0$. 
\end{center}

How can we take each element of the family to be normalizable yet have them converge to the (non-normalizable) residual diff? The idea is to consider a class of perturbations that give a good approximation of the residual diffeomorphism within some finite region of spacetime and send the size of that region to infinity (in the open case) or to the size of the universe itself (in the closed case). For concreteness, let us concentrate on the dilation residual diff (``Res. Diff'') in an open universe. We can take

\begin{eqnarray}\label{formal}
\zeta_{(\alpha)}(t = 0, \v{x}) & = & \zeta_{\text{Res. Diff}}(t=0, \v{x}) \exp \left( - \alpha F(x) \right), \\
\dot{\zeta}_{(\alpha)}(t = 0, \v{x}) & = & \dot{\zeta}_{\text{Res. Diff}}(t=0, \v{x})\exp \left(  - \alpha F(x)  \right),
\end{eqnarray}
where $F(x)$ is a function that increases with $ x $ sufficiently fast for each perturbation to be normalizable. Note that we need to specify the field as well as its time derivative, since the evolution equation for scalars, (\ref{eq:ScalarTimeDep}), is second order in time. It is likely that such a non-monochromatic adiabatic mode would not have the same relation between $  \zeta(t=0, \v{x}) $ and $  \dot\zeta(t=0, \v{x}) $ as the physical pertubation coming from a Bunch-Davies initial state, making these non-monochromatic modes less useful in practice.
 
We make a technical assumption that the Cauchy problem for linear perturbations is well-posed so that the solution must depend continuously on the initial conditions (although this assumption is not strictly necessary if we take different physical field profiles). The physical modes converge pointwise to the residual diff on the initial time slice, so by continuity of solutions they must also converge pointwise to the gauge mode in the entire spacetime region for which solutions exist (a natural assumption is that solutions do exist in the entire spacetime, ie. linear evolution of smooth initial conditions doesn't lead to singularities). This completes the construction of a family of scalar physical modes that become adiabatic in the limit $\alpha \to 0$. The argument is similar in the case of tensor modes and is valid in both open and closed universes. We can also conclude that on any finite (but otherwise arbitrary) patch of spacetime, convergence must be uniform.

 
\section{Soft theorems}\label{sec:SoftThms}


In this section, we start with a brief review of soft theorems in a flat universe to highlight the difficulties in generalizing the usual derivation to a curved universe. We then point out that, due to a gap between the Laplacian eigenvalues of residual diffeomorphisms and physical modes, the soft limit cannot be constrained  in the usual way at order $O(K)$. We also briefly discuss some new non-standard soft theorems of a more formal nature. Later on, in Sec. \ref{sec:SqueezedLimits} and Sec. \ref{sec:ObsSignatures}, we will directly compute curvature corrections to correlators, which will confirm the findings of this section.


\subsection{Flat universe}

Residual diffs reflect the underlying symmetries of the gravitational theory and therefore can be used to derive soft theorems for primordial correlators. A generic argument can be constructed as follows. Consider an $n+1$-point function where one of the modes is close to a residual diff, such that its dominant time dependence matches that of the residual diff. We can write
\begin{align}
\langle \zeta_{\v{q}} \zeta_{\v{k}_1} \ldots \zeta_{\v{k}_n} \rangle & \sim \int \frac{d^3 q'}{(2 \pi)^3} \langle \zeta_{\v{q}} \zeta_{\v{q'}} \rangle \frac{\delta}{\delta \zeta_{\v{q'}}} \langle \zeta_{\v{k}_1} \ldots \zeta_{\v{k}_n} \rangle_{\zeta_{\v{q'}}}+{\cal O}(\zeta_{\v{q}}^{3})\\ \nn
&=P_{\zeta}(q) \frac{\delta}{\delta \zeta_{AM}} \langle \zeta_{\v{k}_1} \ldots \zeta_{\v{k}_n} \rangle_{\zeta_{AM}}+{\cal O}(\zeta_{\v{q}}^{3},q^2/k^2,q^{2}/(aH)^2 ) .
\end{align}

In the final step we used the conservation of momentum and the fact that the soft mode resembles the residual diff (up to corrections of order $q^2/k^2$ and  $q^{2}/(aH)^2$). The effect of a residual diff on the short modes is precisely a change of coordinates $x^{\mu}\to x^{\mu}+\epsilon^\mu$:
\begin{equation}
\delta_{\zeta_{GM}} \langle \zeta_{\v{k}_1} \ldots \zeta_{\v{k}_n} \rangle_{\zeta_{GM}} = \sum\limits_{i=1}^n \langle \zeta_{\v{k}_1} \ldots ( \delta_{\epsilon} \zeta_{\v{k}_i} )  \ldots \zeta_{\v{k}_n} \rangle.
\end{equation}
Then the soft theorem takes the form
\begin{align}
\lim_{q\to 0}\langle \zeta_{\v{q}} \zeta_{\v{k}_1} \ldots \zeta_{\v{k}_n} \rangle \sim P_{\zeta}(q) \sum\limits_{i=1}^n \langle \zeta_{\v{k}_1} \ldots ( \delta_{\epsilon} \zeta_{\v{k}_i} ) \ldots \zeta_{\v{k}_n} \rangle + {\cal O}(q^2/k_n^2,q^{2}/(aH)^2) \,.
\end{align}

\subsection{Absence of monochromatic soft theorems}

In Section \ref{sec:AM}, we have established that in curved universes monochromatic physical modes are always separated from residual diffs by some finite gap that is proportional to $  K $. Thus, a consistency relation in which the soft mode is monochromatic fails to capture the effect of a diff by a discrete amount; a difference of order $\O(K/k_s^2)$ is always present between physical modes and residual diffs. Of course this difference can be very small if curvature is very small, and that limit indeed reproduces the flat space results. But already at linear order in $  K $ one finds violations of the flat-universe soft theorems. We conclude that soft theorems of the usual form do not exist in curved universes. This conclusion applies to both scalar and tensor soft theorems in both open and closed universes. 

We can explicitly show where the $\O(K/k_s^2)$ errors originate in the derivation. Focusing (for concreteness) on the open universe, scalar case, we obtain - in terms of open harmonics:
\begin{align} \nn
\langle \zeta_{q00} \zeta_{k_{s}lm} \zeta_{k'_{s}l'm'} \rangle & = \int dq' \langle \zeta_{q00} \zeta_{q'00} \frac{\delta}{\delta \zeta_{q'00}}  \langle \zeta_{k_{s}lm} \zeta_{k'_{s}l'm'} \rangle_{\zeta} \rangle \\
& \approx 
\int dq' \langle \zeta_{q00} \zeta_{q'00} \rangle \frac{\delta}{\delta \epsilon}
 \langle  \zeta_{k_{s}lm} \zeta_{k'_{s}l'm'} \rangle + \O \left( \frac{q^2 - 3K}{k_{s}^2} \right) \langle \zeta_{k_{s}lm} \zeta_{k'_{s}l'm'} \rangle . 
\end{align}
The error in approximating a physical mode $\zeta_{q'00}$ with a monochromatic residual diff as done in the second line comes from two effects. First, even at some constant time we know that the eigenvalues of the Laplacian must differ at order $\O\left( q^2 - 3K \right) $. Second, by the differential equation that each perturbation satisfies, see for example \eqref{eq:ScalarTimeDep}, this difference in $  \nabla^{2} $ leads to a time dependence that is also different at order $  \O(q^{2}-3K) $, since this quantity vanishes for the residual scalar diff but it does not for the physical mode. A short mode with momentum $  k_{s} $ and associated length scale $  x_{s}=k_{s}^{-1} $ feels this difference as its evolves before freezing out. The relevant dimensionless quantities that estimate this error are then
\begin{align}
\O \left(( q^{2}-3K)x_{s}^{2} \right) \simeq \O \left(  \frac{q^{2}-3K}{k_{s}^{2}} \right) \geq \O \left( \frac{|K|}{k_{s}^{2}} \right) \,.
\end{align} 

Note that the error does not contain an $\O ( |K|/ q_{l}^{2} )$ term. Thus the leading order behaviour of a model dependent effect in the squeezed bispectrum will be $\O ( |K|/ k_{s}^{2} )$, as we will show explicitly in Sec. \ref{sec:SqueezedLimits}.

\subsection{Formal soft theorems and adiabatic modes}

We know from the discussion around \eqref{formal} that in curved universes physical perturbations can be approximated arbitrarily well by residual diffs, even though such perturbations cannot be monochromatic and do not come from a Bunch-Davies initial condition. Consider then any family of physical scalar perturbations $\zeta_{(\alpha)}$ that are adiabatic in the limit $\alpha \to 0$. The strategy is to put the $\zeta_{(\alpha)}$ modes on the left hand side of a consistency relation in place of the usual soft mode. 

Let us assume that for each $\alpha$, there exists a basis $\{ \zeta_{\v{n}}\}$ (its elements labelled by $\v{n}$) of \textit{pairwise uncorrelated modes} containing $\zeta_{(\alpha)}$ itself. Then we have
\begin{equation}
\langle \zeta_{(\alpha)} \zeta_{klm} \zeta_{k'l'm'} \rangle {\sim} \int d \v{n} \langle \zeta_{(\alpha)} \zeta_{\v{n}}  \rangle \frac{\delta}{\delta \zeta_{\v{n}}} \langle  \zeta_{klm} \zeta_{k'l'm'} \rangle_{\zeta_{\v{n}}} \sim \langle \zeta_{(\alpha)}^2 \rangle' \frac{\delta}{\delta \zeta_{\text{Res. diff}}} \langle  \zeta_{klm} \zeta_{k'l'm'} \rangle_{\zeta_{\text{Res. diff}}},
\label{eq:SoftTheorem}
 \end{equation}
up to corrections that vanish in the limit $\alpha \to 0$, in which the soft perturbation approaches the residual diff.

A few remarks are in place. The primed $2$-pt function $ \langle \zeta_{(\alpha)}^2 \rangle'$ depends on the chosen basis $\{ \zeta_{\v{n}}\}$ and it will be generally difficult to compute. Similarly, we do not have a closed expression for the action of a diff on the short modes. Finally, $O(K)$ effects in the consistency relation are correctly captured only when the soft mode $\zeta_{(\alpha)}$ resembles the adiabatic mode on scales comparable to the curvature scale, in which case $ \langle \zeta_{(\alpha)}^2 \rangle'$ is a superhorizon quantity inaccessible to local observers. Despite all these limitations, the above formulation of a soft theorem is formally valid to an arbitrary accuracy in the limit $\alpha \to 0$, despite the existence of a gap in the momenta of residual diffs and physical modes. We thus demonstrated the possibility of extending soft theorems to universes with nonvanishing spatial curvature, albeit in a formal sense.


\section{Curvature corrections to the power spectrum and bispectrum}\label{sec:SqueezedLimits}



The primary focus of this section is to investigate the dominant effect of curvature on the power spectrum and the bispectrum
\begin{align}
\langle \zeta_{\v{k_1}} \zeta_{\v{k_2}}   \rangle &= (2 \pi)^3 P(k_{1}) \delta^{(3)}\left( \v{k}_{1}+\v{k}_{2}\right) \,,\\
\langle \zeta_{\v{k_1}} \zeta_{\v{k_2}} \zeta_{\v{k_3}} \rangle &= (2 \pi)^3 B(k_1, k_2, k_3) \delta^{(3)}\left( 
\bfk_1+\bfk_2+\bfk_3 \right) \,.
\end{align}
Whenever the scale associated with curvature is much larger than length scales relevant to observations, the physical effects of the former can be captured by an isotropic perturbation in a flat universe \cite{Baldauf:2011}. 
To see this, let expand the metric around the origin of coordinates, or equivalently to linear order in $  K $:
\begin{align}
g_{ij}&= \frac {a_{K}(t)^{2}\delta_{ij}} { \left( 1 + \frac 1 4 K \bfx^2 \right)^{2}}\simeq a_{K}^{2}\delta_{ij}\left( 1-\frac{1}{2} K \bfx^2 \right)+ {\cal O}((K\textbf{x}^2)^2)\\
&=a^{2}\delta_{ij}\left[ 1+2 \frac{K}{a} \left( \frac{\partial a_{K}}{\partial K} \right)_{K=0}-\frac{1}{2} K \bfx^2 \right]+ {\cal O}((K\textbf{x}^2)^2\,,
\end{align}
where $  a_{K} $ is the solution of the Friedmann equation in presence of spatial curvature while $  a $ is the solution for a flat FLRW metric. We can re-interpret this metric in terms of the flat space metric with a curvature perturbation
\begin{align}
g_{ij}=a^{2}_{flat} \delta_{ij}(1+2 \tilde \zeta_{B})\,.
\end{align}
There are two possible ways to do this\footnote{To avoid confusion, \textit{specific field configurations} will be marked by a tilde, as in $\tilde{\zeta}_B$.}: 
\begin{align}
\text{Option 1:}& &a_{\text{flat}}&= a_{K} & \tilde \zeta_{B}& = -\frac{1}{4} K \bfx^2  \,,\label{eq:BackgroundCurvature}\\
\text{Option 2:}& &a_{\text{flat}}&= a & \tilde \zeta_{B} & = \zeta_{K}(t)-\frac{1}{4} K \bfx^2  \,,
\end{align}
where we defined
\be\label{zetaK}
\zeta_K(t) \equiv \frac{K}{a(t)}\left[ \frac{\partial a_K(t)}{\partial K}\right ]_{K=0}\,.
\ee 
Option 1 has the advantage that $  \tilde \zeta_{B} $ is simpler, but it has the awkward feature that the curvature effects are partly in perturbations and partly in the background quantities, which mixes the perturbative expansion with the $  K $ expansion. Option 2 instead has the advantage that $  \tilde \zeta_{B} $ is precisely a solution of the Einstein equations in a flat universe, but the drawback is that it contains one more term. 
In the following we will find it more convenient to work with Option 1. In words, it says that the spherical perturbation in \eqref{eq:BackgroundCurvature} is locally indistinguishable from a mean curvature $K$ superimposed with a change in the scale factor or equivalently in the Hubble parameter,\footnote{$H$ is the Hubble parameter in the absence of curvature, while $H_K$ is the Hubble parameter when curvature is present.} 
\begin{equation}
H_K \to H\, \left[ 1+\dfrac{K}{2a^2(t)H^2} \right] \,.
\end{equation}
\noindent
(This is valid for attractor scenarios, where superhorizon modes are approximately frozen.) Thus, if we knew the $(n+1)$-point function $\langle \zeta_{\v{q_u}} \zeta_{\v{k_1}}\ldots \zeta_{\v{k_n}} \rangle$ in the regime $q_u \ll k_i$, we could recover the leading order effect of curvature on the $n$-point function. We expect that order $K$ corrections to the power spectrum and to the bispectrum obtained through this method are in no way constrained by soft theorems. This is because $\O(K)$ in the $n$-point function corresponds to $\O(q_u^2)$ in the  $(n+1)$-pt function, which is already a model-dependent effect. Hence, $\O(K)$ corrections will depend on the details of inflationary theory, and might be enhanced even in single field scenarios. 

In this section, we quote the result of \cite{Creminelli:2013cga} for the curvature correction to the \textit{power spectrum} in the framework of the EFT of inflation with small speed of sound. Then we study $\O(K)$ corrections to the \textit{bispectrum} in two cases: canonical, single field, slow roll inflation as well as the EFT of inflation. Our argument follows closely the method of \cite{Creminelli:2013cga}, and we compare our results to the $\O(K)$ correction to the power spectrum derived therein. It must be noted that we work with Fourier modes defined with respect to the coordinates and metric of the \textit{flat} background, that is, when the curvature is treated as a separate perturbation introduced on top of the flat reference space.

One caveat of our analysis is that we will be assuming a Bunch-Davies initial state in calculating correlators. In a curved universe, the Euclidean vacuum instead gives a modified initial state (see for example \cite{Sasaki:1994yt,Yamamoto:1995sw,Yamamoto:1996qq,Tanaka:1997kq}). The difference between these two initial states is non-perturbative in $  K $ and therefore cannot be captured by the arguments in this section. We have performed a preliminary investigation of the relative importance of the initial state modification in canonical inflation as compared with the perturbative corrections using the analytical results of \cite{Garriga:1998he}. We found that the non-perturbative terms give an effect that is numerically negligible in the final primordial power spectrum for the parameters relevant to this work. This is to be expected as the deviation from Bunch-Davies is non-perturbative in curvature, and consistency forces us to remain in the perturbative regime. Because of this we will systematically neglect these corrections.


\subsection{The power spectrum}\label{ssec:SPS}

The effect of spatial curvature to the power spectrum can be calculated at linear order in $  K $ by summing up two physical effects: the presence of the ultra-long mode $\tilde{\zeta}_B$ and a change in the Hubble parameter. 

The effect of the ultra-long mode was computed to leading order in $K$ in \cite{Creminelli:2013cga} by considering the $3$-pt function (in the EFT of inflation) in the squeezed limit, as explained in the previous subsection. The result is
\begin{equation}
\label{paolo}
\Delta_{\tilde \zeta_{B}} P_{K}(k) = - P_{\text{flat}}(k) \frac{19 + 6 c_3}{8 c_s^2} \frac{K}{k^2} .
\end{equation}
The effect of the change in Hubble parameter can be found from the familiar relation $P_{\text{flat}}(k) =H_*^2/(4 \epsilon k^3)$. To leading order,
\begin{equation}\label{H}
\Delta_{H} P_{K}(k) = P_{\text{flat}}(k)  \frac{K}{k^2} \,,
\end{equation}
so this effect is subdominant for small $c_s$.

In addition to the above there is yet another effect of a more geometrical nature. When observations of the sky are performed, some assumptions have to be made regarding the connection between position space and Fourier space power spectra. While this correspondence is unambiguous in flat space (up to constants of proportionality), in a spatially curved universe there exist multiple conventions that could lead to slightly different Fourier space results.\footnote{As an example, consider \textit{flat Fourier modes}, defined as the Fourier transform of perturbations on a flat reference background:
\begin{equation*}
\zeta^{\text{flat}}_{\v{k}} = \int d^3 x e^{ - i \v{k} \cdot \v{x}} \zeta^{\text{flat}}(x) 
\end{equation*}
and compare this with the Fourier transform of perturbations on top of a curved background, with respect to the coordinate system defined from the apparent distance to objects in a curved universe,
\begin{equation*}
\zeta_{\v{k}} = \int d^3 r e^{ - i \v{k} \cdot \v{r}} \zeta(r) .
\end{equation*}
Because $\zeta(r) \not\equiv \zeta^{\text{flat}}(x = r)$ (rather, there is a discrepancy of order $Kx^2$), the two Fourier transforms are not equivalent.} All the power spectra must give the same flat space limit and their ratio is a purely geometrical quantity that can only depend on $K$ and the momentum, not on the physical quantities such as the EFT parameters and in particular $  c_{s} $. Thus, in general the ``geometrical'' effects contribute
\begin{equation}\label{geo}
\Delta_{\text{geom.}}P_{K}(k) = P_{\text{flat}}(k)  \O(K/k^2)\,,
\end{equation}
and are again subdominant for small $c_s$. We will neglect both \eqref{H} and \eqref{geo} in the analysis in Sec. \ref{sec:ObsSignatures}.

In conclusion, to leading order in $c_s \to 0$ the only contributing term is $  \delta_{\tilde \zeta_{G}} $. So to linear order in $K$ the power spectrum becomes
\begin{equation}
P_{K}(k) = A_s k^{(n_s - 1 ) -3} \left( 1 - \frac{19 + 6 c_3}{8 c_s^2} \frac{K}{k^2} \right) +\O(K c_{s}^{0}) .
\label{eq:PScurved}
\end{equation}
\begin{figure}
\begin{center}
\includegraphics[scale=0.12]{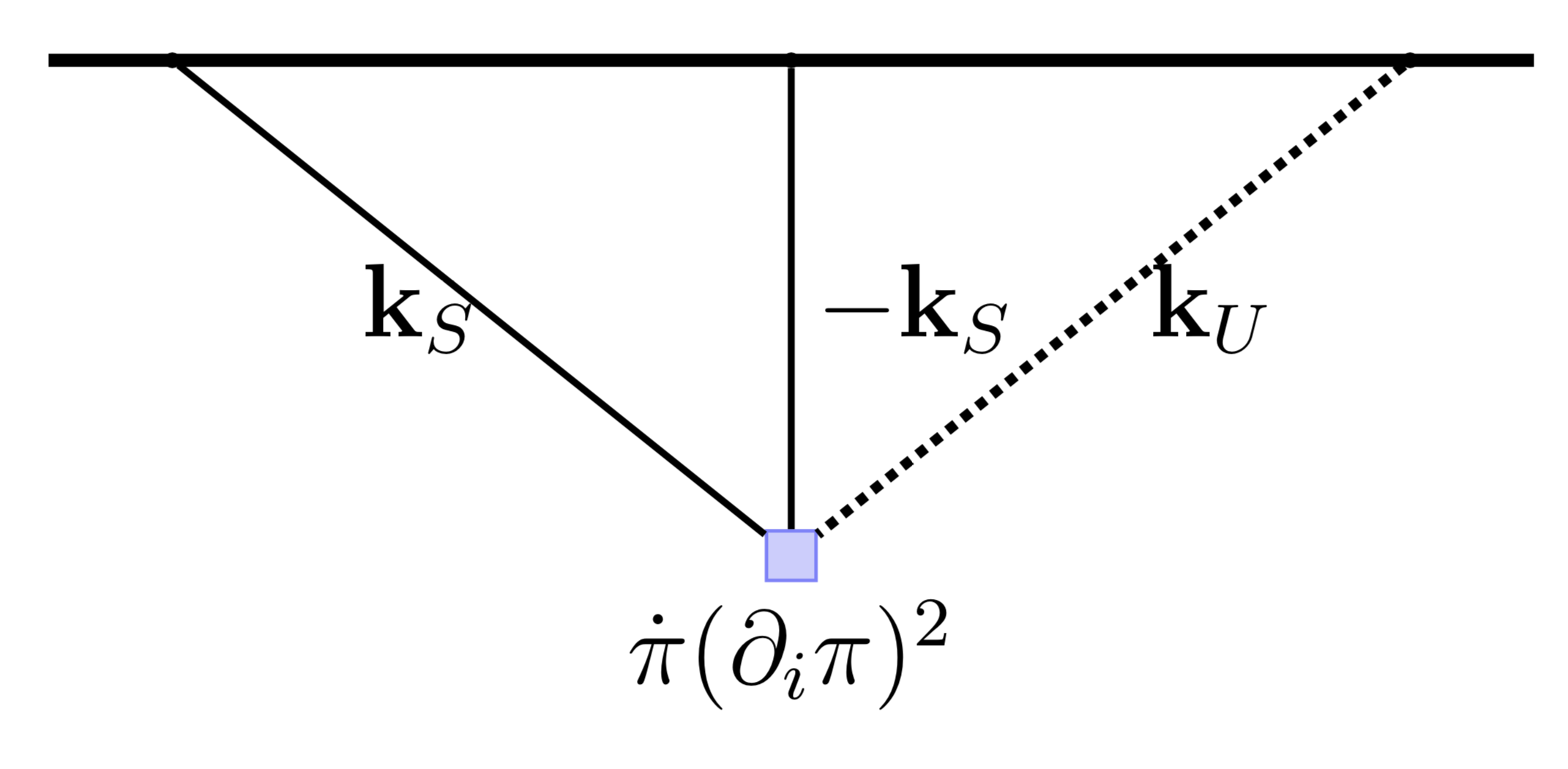}
\includegraphics[scale=0.125]{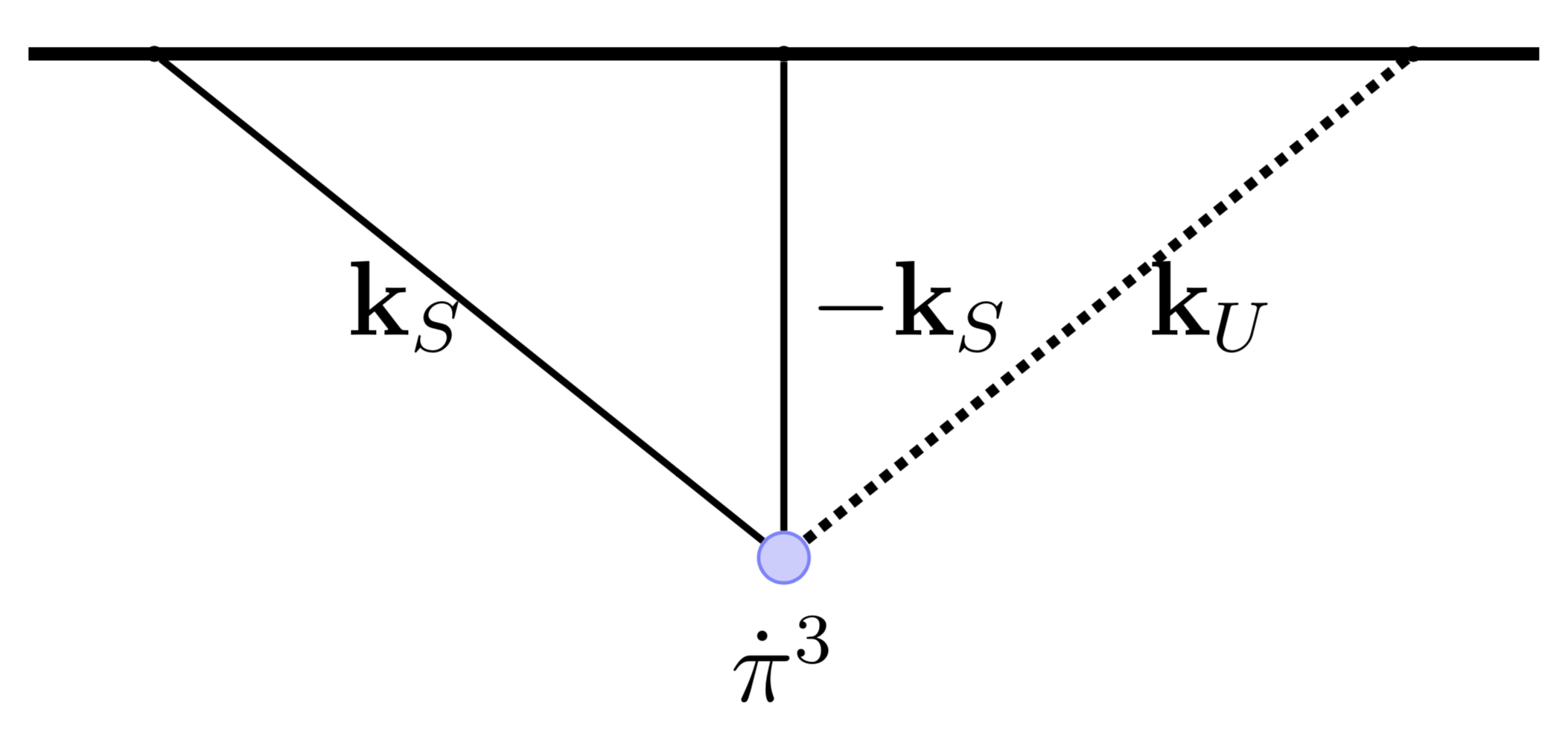}
\end{center}
\caption{Three-point function diagrams for EFT of inflation. The ultra-long mode $k_U$ mocks the effect of spatial curvature. }
\end{figure}


\subsection{Background curvature argument}
\label{bgcurv}
Let us assume that $K>0$, and later we can analytically continue our results to $K<0$. Recall from the discussion in the beginning of this section that we can trade curvature for the following spherically-symmetric perturbation around flat space,
\begin{equation}
\label{bg}
\tilde{\zeta}_B(t,\v{x}) = - \frac{1}{4}K \textbf{x}^2 + {\cal O}((K\textbf{x}^2)^2)\,,
\end{equation}
 plus a modification in the scale factor, i.e. $  a_{\text{flat}}= a_{K} $. This implies that, for example, a three-point function of $\zeta$ in a curved universe equals the same three-point function in the background of $\tilde{\zeta}_B$ in a flat universe (with a modified scale factor $a_K(t)$). \footnote{Here we assume that the geometric effects stemming from the curvature of spatial slices are negligible with respect to the enhanced effects originating from non-gaussianity, e.g. we still expand $\zeta$ in terms of plane waves and ignore that the sum of spatial momenta does not vanish.}
, i.e. 
\begin{equation}\label{15}
\langle \zeta_{\v{k}_1} \zeta_{\v{k}_2} \zeta_{\v{k}_3} \rangle_{K, H_K} =
\langle \zeta_{\v{q}} \zeta_{\v{k}} \zeta_{-\v{k}} \rangle_{\tilde{\zeta}_B, H_{K}}+ {\cal O}((K\textbf{x}^2)^2),
\end{equation}
where correlators are assumed to be taken in flat space unless otherwise specified by the label ``$  K $''. In the following we will keep implicit in all formulae that the scale factor should be $ a_{K} $ and we will come back to this issue at the end of this section. 
The configuration given by $\tilde{\zeta}_B$ can be mimicked by the long mode limit of the superposition of three orthonormal plane waves (all having the same momentum magnitude $|p_a|=\sqrt{K/2}$ for $a=1,2,3$, and with $\hat{p}_a.\hat{p}_b=\delta_{ab}$) subtracted with an inconsequential constant,  
\be
\label{zetaB0}
\tilde{\zeta}_B(t,\bfx)=\sum\limits_{a=1}^{3} \left[ \cos(\textbf{p}_{a}\cdot\bfx) -1\right]+ {\cal O}((K\textbf{x}^2)^2)\,.
\ee
It can be easily seen that the expression on the RHS starts at order $p^2$ and that it coincides with \eqref{bg}.
The three-point function on a slowly varying background $\zeta_B$ (inclusive of $\tilde{\zeta}_B$) is given by 
\begin{eqnarray}\label{17}
\langle \zeta_{\v{k}_1} \zeta_{\v{k}_2} \zeta_{\v{k}_3} \rangle_{\zeta_B} & \cong & \langle \zeta_{\v{k}_1} \zeta_{\v{k}_2} \zeta_{\v{k}_3} \rangle + \zeta_B\left[\frac{\partial}{\partial \zeta_B}   \langle \zeta_{\v{k}_1} \zeta_{\v{k}_2} \zeta_{\v{k}_3} \rangle_{\tilde{\zeta}_B} \right]_{\zeta_B = 0}\\ \nonumber
&&+\partial_i\zeta_B\left[\frac{\partial}{\partial (\partial_i\zeta_B)}   \langle \zeta_{\v{k}_1} \zeta_{\v{k}_2} \zeta_{\v{k}_3} \rangle_{\zeta_B} \right]_{\zeta_B = 0}\\ \nonumber 
&&+\partial_i\partial_j\zeta_B\left[\frac{\partial}{\partial (\partial_i\partial_j\zeta_B)}   \langle \zeta_{\v{k}_1} \zeta_{\v{k}_2} \zeta_{\v{k}_3} \rangle_{\zeta_B} \right]_{\zeta_B = 0}+\dots \,.
\end{eqnarray}
Therefore, 
\begin{eqnarray}
\langle \zeta_{\v{k}_1} \zeta_{\v{k}_2} \zeta_{\v{k}_3} \rangle_{\tilde{\zeta}_B}& = &  \langle \zeta_{\v{k}_1} \zeta_{\v{k}_2} \zeta_{\v{k}_3} \rangle - \frac{3}{2}K \delta_{ij} \left[\frac{\delta}{\delta (\partial_i\partial_j \zeta_B)}  \langle \zeta_{\v{k}_1}  \zeta_{\v{k}_2} \zeta_{\v{k}_3} \rangle_{\zeta_B} \right]_{\zeta_B = 0}\,.
\end{eqnarray}
For simplicity we drop $\left[... \right]_{\zeta_B=0}$ in the remainder. In order to find the last term in the expression above, we consider the correlation between an ultra-long monochromatic mode, $\tilde{\zeta}_{\v{p}}(t,\bfx)=\zeta_{\v{p}}(t)\,\exp(i\textbf{p}\cdot\textbf{x})$, and three short modes. Up to leading order in the gradients of the ultra-long mode, this trispectrum can be simplified into
\begin{eqnarray}
\langle \zeta_{\v{p}} \zeta^3 \rangle & = &P_\zeta(p)\, \left[ \frac{\partial }{ \partial \zeta_B} \langle \zeta^3 \rangle_{\zeta_B} + i p^i.\frac{\partial}{\partial(\partial_i \zeta_{B})} \langle \zeta^3 \rangle_{\zeta_B}-p^i\,p^j \frac{\partial}{\partial ( \partial_i\partial_j\,\zeta_B)} \langle \zeta^3 \rangle_{\zeta_B}+{\cal O}(p^3)  \right] \\ \nonumber
& = & \text{S.T.} +P_\zeta(p) \left[-p^i\,p^j \frac{\partial}{\partial ( \partial_i\partial_j\,\zeta_B)} \langle \zeta^3 \rangle_{\zeta_B}+{\cal O}(p^3)  \right]\,,
\end{eqnarray}
%
where ``S.T.'' stands for the ${\cal O}(p^0)$ and ${\cal O}(p)$ parts of the correlator, which are fixed by soft theorems \cite{Hinterbichler:2012nm}. Summing over three orthogonal $\textbf{p}$'s one finds
\footnote{Since we are interested in ${\cal O}(K)$ corrections, it is allowed to send $p\to 0$ although we had assumed $p=\sqrt{K/2}$.}
\begin{eqnarray}
\sum\limits_{i=1}^{3}\,\langle \zeta_{\v{p}_i} \zeta^3 \rangle =\text{S.T.} +P_\zeta(p) \delta_{ij}\left[-p^2\, \frac{\delta}{\delta ( \partial_i\partial_j\,\zeta_B)} \langle \zeta^3 \rangle_{\zeta_B}+{\cal O}(p^3)  \right]  \,.
\end{eqnarray}
Equivalently, we can write
\begin{equation}
\delta_{ij}\frac{\delta}{\delta ( \partial_i\partial_j\,\zeta_B)} \langle \zeta^3 \rangle_{\zeta_B}= - \lim\limits_{p\to 0}\frac{1}{2} \frac{\partial^{2}}{\partial p^{2}}\sum\limits_{i=1}^3\left[ P(p)^{-1}  \langle  \zeta_{\v{p}_i} \zeta^3 \rangle  \right].\label{22}
\end{equation}
Finally, by putting together \eqref{15}, \eqref{17} and \eqref{22} we find
\begin{equation}
\boxed{
\langle \zeta_{\v{k}_1} \zeta_{\v{k}_2} \zeta_{\v{k}_3} \rangle_{K} \cong \langle \zeta_{\v{k}_1} \zeta_{\v{k}_2} \zeta_{\v{k}_3} \rangle_{0}  + \frac{3}{2} K \lim\limits_{p \to 0} \frac{1}{2} \frac{\partial^{2}}{\partial p^{2}}\sum\limits_{i=1}^3  P(p)^{-1}  \left[ \langle  \zeta_{\v{p}_i} \zeta_{\v{k}_1} \zeta_{\v{k}_2} \zeta_{\v{k}_3} \rangle  \right]\,.}\
\label{eq:BCA}
\end{equation}


\subsection{Effects of curvature in the bispectrum: the EFT of inflation}
The effective field theory (EFT) of inflation \cite{Creminelli:2006xe,Cheung:2007st} is the general theory of single-field fluctuations around any FLRW spacetime.
In place of the covariant scalar field $\phi$, the action is expressed in terms of the Goldstone mode of the broken time translation, defined through $\phi(t, \v{x})=\bar{\phi}(t+\pi)$ (where $\bar{\phi}$ is the background of the scalar field). Asking $\pi$ to non-linearly realize the temporal diffs and linearly realize the spatial ones fixes the dynamics up to some arbitrary time dependent functions. 

At sufficiently short distances, the physics of $\pi$ decouples from the metric perturbations. In this paper, we will assume that the time dependence of all background quantities is sufficiently slow that we can approximate them as constant\footnote{Notice that this is not the same as assuming a shift symmetry, as discussed in details in \cite{Finelli:2018upr}.}. The EFT action in the decoupling limit up to cubic order in slow-roll corrections
and up to quartic order in the field becomes
\begin{eqnarray}
S_{\pi} = \int d^4 x \sqrt{-g}\, \dfrac{\epsilon H^2 M_p^2}{c_s^2} &&\left[  \left( \dot{\pi}^2 - c_s^2 \dfrac{(\nabla \pi)^2}{a^2} \right) + C_{\scriptscriptstyle\dot{\pi}(\nabla \pi)^2}\dot{\pi}\dfrac{(\nabla \pi)^2}{a^2}\right. \\ \nonumber
&& \left. + C_{\scriptscriptstyle\dot{\pi}^3}\dot{\pi}^3 +C_{\scriptscriptstyle(\nabla \pi)^4}\dfrac{1}{a^4}(\nabla \pi)^4 +C_{\scriptscriptstyle \dot{\pi}^2(\nabla\pi)^2}\dfrac{1}{a^2}\dot{\pi}^2(\nabla\pi)^2+C_{\dot{\pi}^4}\dot{\pi}^4\right]\,.\label{eq:EFTaction}
\end{eqnarray}
As for the cubic operators, the coefficient $C_{\scriptscriptstyle \dot{\pi}(\nabla \pi)^2}$ is entirely determined by demanding the non-linear realization of Lorentz boosts
\begin{equation}
C_{\scriptscriptstyle \dot{\pi}(\nabla \pi)^2}=c_s^2-1\,, 
\end{equation}
while $C_{\scriptscriptstyle\dot{\pi}^3}$ is a free time-dependent function and is conventionally parametrised as
\begin{equation}
C_{\scriptscriptstyle\dot{\pi}^3}=(1-c_s^2)\left( 1 + \frac{2}{3} \frac{c_3}{c_s^2} \right).
\end{equation}
As for the quartic interactions, invariance under boosts relates two of the coefficients to the cubic operators, namely
\be
C_{\scriptscriptstyle (\nabla\pi)^4}=-C_{\scriptscriptstyle \dot{\pi}(\nabla \pi)^2}\,,\qquad\qquad C_{\scriptscriptstyle \dot{\pi}^2(\nabla\pi)^2}=-\dfrac{3}{2}(C_{\scriptscriptstyle \dot{\pi}(\nabla \pi)^2}-C_{\scriptscriptstyle \dot{\pi}^3})\,,
\ee
whereas, $C_{\dot{\pi}^4}$ is a free coefficient. Inasmuch as  the $\dot{\pi}^4$ operator is unconstrained by the constraints on the bispectrum, it is allowed to be much bigger than $\dot{\pi}^2(\partial_i\pi)^2$ and $(\partial_i\pi)^4$ \cite{Smith:2015uia} . For this reason, in the remainder we only keep the contribution of $\dot{\pi}^4$ to the trispectrum. 

Now we turn to computing the bispectrum of $\zeta$ in the EFT of inflation to leading order in $K$ and slow-roll corrections by using the method explained in \ref{bgcurv}. The trispectrum generated by operators in \eqref{eq:EFTaction} was calculated in \cite{Chen:2009se}. The cubic operators contribute to the trispectrum via four types of exchange diagrams depicted in Figure \ref{feynmans}. Below, we separately give the bispectrum generated by individual Feynmann diagrams in Figure \ref{feynmans}, namely 
$B_{\dot{\pi}^3\, \dot{\pi}^3}$, corresponding to the exchange diagram with two $\dot{\pi}^3$ vertices, 
$B_{\dot{\pi}^3\, \dot{\pi}(\nabla \pi)^2}$ for the diagram with one $\dot{\pi}^3$ and one $\dot{\pi}(\nabla\pi)^2$ vertex, and finally $B_{\dot{\pi}(\nabla\pi)^2\, \dot{\pi}(\nabla\pi)^2}$ representing the diagram with two $\dot{\pi}(\nabla\pi)^2$ vertices. Since the final answer is symmetric under permutations of momenta, we give the expressions in terms of the elementary symmetric polynomials, defined by
\be
e_1=\sum\limits_{i=1}^3\, k_i\,,\qquad e_2=\sum\limits_{i<j}\, k_i\,k_j\,,\qquad e_3=\prod\limits_{i=1}^3\,k_i\,.
\ee
We find
\begin{eqnarray}\label{uno}
B_{\dot{\pi}^3\, \dot{\pi}^3} &=& 18  \left(\dfrac{H^2}{4\epsilon c_s M_p^2}\right)^2\,\left(\prod\limits_{i=1}^3\,\dfrac{1}{2k_i^3}\right)\,\dfrac{K}{c_s^4\,e_1^5}\,(-1+c_s^2)^2 (2c_3+3c_s^2)^2\,\\ \nonumber 
&&\qquad \times \left(-2 e_3 e_1^3+e_2^2 e_1^2+3 e_2 e_3 e_1+76 e_3^2\right)\,,\\ \nonumber
&&\\ 
B_{\dot{\pi}^3\, \dot{\pi}(\nabla\pi)^2}&=& \frac{3}{2} \left(\dfrac{H^2}{4\epsilon c_s M_p^2}\right)^2\,\label{due}\left(\prod\limits_{i=1}^3\,\dfrac{1}{2k_i^3}\right)\,\dfrac{\,K}{c_s^4\,e_3^2\,e_1^5} (-1+c_s^2)^2 (2c_3 + 3c_s^2)\times\\ \nonumber
&& \left[  6 e_3 e_1^9-3 e_2^2 e_1^8-18 e_2 e_3 e_1^7+9 \left(e_2^3+5 e_3^2\right) e_1^6-57 e_2^2 e_3 e_1^5-1184 e_3^4\right. \\ \nonumber
&&+12 e_2 \left(e_2^3+e_3^2\right) e_1^4 
 \left. +4 \left(3
   e_2^3 e_3-107 e_3^3\right) e_1^3+232 e_2^2 e_3^2 e_1^2+672 e_2 e_3^3 e_1\right ]\,,\\ \nonumber
  && \\
B_{\dot{\pi}(\nabla\pi)^2\, \dot{\pi}(\nabla\pi)^2}&=& - \frac{3}{2} \left(\dfrac{H^2}{4\epsilon c_s M_p^2}\right)^2\,\label{tre}\left(\prod\limits_{i=1}^3\,\dfrac{1}{2k_i^3}\right)\,\dfrac{\,K}{c_s^4\,e_1^5\,e_3^2}\times\\ \nonumber
&& \left[-38 e_3 e_1^9+19 e_2^2 e_1^8+114 e_2 e_3 e_1^7-3 (19 e_2^3+135 e_3^2) e_1^6 \right. \\ \nonumber
&&\left. +361 e_2^2 e_3 e_1^5+(44 e_2 e_3^2-76 e_2^4)+452 e_2^2 e_3^2 e_1^2
   e_1^4\right. \\ \nonumber
&&\left. -4 (19 e_3 e_2^3+253 e_3^3) e_1^3+1508 e_2 e_3^3 e_1-2256 e_3^4\right ]\,.
\end{eqnarray}
The only contact term that we consider is generated via the operator $\dot{\pi}^4$, and the resulting bispectrum is
\footnote{Notice that in the interaction Hamiltonian the coefficient of the $\dot{\pi}^4$ term  differs from the one in the original Lagrangian due to the correction that the conjugate momentum of $\pi$ receives from the cubic term $\dot{\pi}^3$.}
\be
B_{\dot{\pi}^4}=3 \times 36 \times 96 \left(\dfrac{H^2}{4\epsilon c_s M_p^2}\right)^2\,\left(\prod\limits_{i=1}^3\,\dfrac{1}{2k_i^3}\right) \,K \left[ \,C_{\dot{\pi}^4}-\dfrac{9}{4}\left(1+\dfrac{2}{3}\dfrac{c_3}{c_s^2}\right)^2 \right]\,\dfrac{e_3^2}{e_1^5}\,.\label{pidot}
\ee


\subsubsection{Squeezed limit in the EFT of inflation}

Let us compute the squeezed limit of the curvature corrections to the bispectrum in a curved universe, in the framework of the EFT of inflation. In particular, we want to find the dominant $\O(K)$ corrections to the bispectrum in the regime $c_s \ll 1$ in terms of the two EFT quantities $c_s$ and $c_3$ that parameterize the 3-point vertices. This can be done by taking the squeezed limit of \eqref{uno}-\eqref{tre}. For a more explicit derivation we can evaluate the trispectrum
\begin{equation}
\langle \zeta_{\v{q_u}} \zeta_{\v{q_l}} \zeta_{\v{k_s}} \zeta_{\v{k_{s'}}}   \rangle\,
\end{equation}
in the double squeezed limit
\begin{align}
q_{u}\ll q_{l}\ll k_{s} \sim k_{s'}\, ,
\end{align}
and use the background curvature argument to find the $\O(K)$ term in the bispectrum. We leave the details of the calculation to Appendix \ref{app:EFTTrispectrum}. In conclusion, we find the following curvature corrections to the squeezed EFT bispectrum:
\begin{align}\label{final2}
\frac{B(q_l, |\v{k}_s - \frac{1}{2}\v{q}_l |, |\v{k}_s + \frac{1}{2}\v{q}_l |)}{P_{\zeta}(q_l)  P_{\zeta}(k_s)} &\sim (1-n_s) +   \frac{c_s^{2}-1}{c_{s}^{2}}  \left[ \left( 2 + \frac{1}{2}c_3 + \frac{3}{4} c_s^2 \right) - \frac{5}{4} ( \hvec{q}_l \cdot \hvec{k}_s )^2 \right]  \frac{q_l^2}{k_s^2}  \\
& +   \frac{3}{2}  c_s^{-4} \left[ \left( \frac{3}{4} c_3^2 + \frac{43}{8}c_3 + \frac{19}{2} \right)   -  \left(\frac{15}{8}c_3 + \frac{95}{16} \right) ( \hvec{q}_l \cdot \hvec{k}_s )^2 \right] \frac{K}{k_s^2},  \nonumber
\end{align}
where the first line is the flat-space result obeying Maldacena's consistency relation and the second line is the curvature correction. 
Notice that there is an interesting relation between the leading-order curvature correction in the squeezed bispectrum and in the power spectrum. If we average the second line of \eqref{final2} over the angle $  \theta $ between $  \hvec{q}_{l} $ and $ \hvec{k}_s  $, the term $  ( \hvec{q}_l \cdot \hvec{k}_s )^2 $ reduces to a factor of $  1/3 $ and we find
\begin{align}\label{relation}
\int d \cos\theta \,\Delta_{\tilde \zeta_{B}} B = 2 \left( \Delta_{\tilde \zeta_{B}} P \right)^{2}\,,
\end{align}
where $  \Delta_{\tilde \zeta_{B}} P $ is the leading-order curvature correction to the power spectrum discussed around \eqref{paolo} and $  \Delta_{\tilde \zeta_{B}} B $ is the second line of \eqref{final2}. This can be understood in various ways. For example, recall that from the wavefunction of the universe $  \psi $, which takes the form
\begin{align}
\psi=\exp \left[  -\sum_{n=2}^{\infty} \int \frac{1}{n!}\psi_{n} \zeta^{n}\right]\,,
\end{align}
we can derive the following expressions for the correlators
\begin{align}
P(k)&=\frac{1}{2\Re \psi_{2}(k)}\,,\\
B(q_{u},k_{s},k_{s'})&=-\frac{1}{4}\frac{\Re \psi_{3}(q_{u},k_{s},k_{s'})}{\Re \psi_{2}(q_{u})\Re \psi_{2}(k_{s})\Re \psi_{2}(k_{s'})}\,,\\
T(\vec q_{u},\vec q_{l},\vec k_{s},\vec k_{s'})&=-\frac{1}{8\Re \psi_{2}(q_{u})\Re \psi_{2}(q_{l})\Re \psi_{2}(k_{s})\Re \psi_{2}(k_{s'})} \times \left[  \Re \psi_{4}(\vec q_{u},\vec q_{l},\vec k_{s},\vec k_{s'}) + \right. \\
&\left.- \frac{\Re \psi_{3}(q_{u},q_{l},k_{I})\Re \psi_{3}(k_{s},k_{s'},k_{I})}{\Re \psi_{2}(k_{I})} + \text{2 perm's}\right]\,,\label{t2b}
\end{align} 
where $  \vec k_{I}= \vec q_{u} + \vec q_{l}$. These expressions are valid for any momenta, but we have chosen the momenta to match the derivation of curvature effects from flat-universe correlators.
Because we found that the flat-universe four-point interaction does not contribute to curvature effects in the squeezed bispectrum, we can neglect $  \psi_{4} $ above and we see that $  T $ is related to $  B^{2} $. Also, only the permutation displayed contributes in the relevant limit $  q_{u}\ll q_{l} \ll k_{s}\sim k_{s}' $ and $  k_{I}\simeq q_{l} $. Following the strategy outlined earlier in this section, we can extract from the flat universe $  T $ the curvature correction to $  B $ in a curved universe by averaging over the direction of $  \vec q_{u} $. Upon doing this, we see that one of the $  \psi_{3} $  on the right-hand side of \eqref{t2b}, which can be traded for $  B $, gets also angle-averaged and becomes $  \Delta_{\tilde \zeta_{B}} P $. To get to \eqref{relation} we need to also angle average over $ \vec q_{l}  $, which transforms the second $  \psi_{3} $ into a second factor of $   \Delta_{\tilde \zeta_{B}} P $.


\subsection{Effects of curvature in the bispectrum: canonical, single-field inflation}

We now study Einstein gravity coupled to a single scalar inflaton field, which gives a simple action
\begin{equation}
S = \frac{1}{2} \int d^4 x \sqrt{-g} \left( R - (\nabla \phi)^2 - 2 V(\phi) \right).
\label{eq:CanAction}
\end{equation}
We would like to compute $\O(K)$ contributions to the bispectrum $\langle \zeta_{\v{q}_l} \zeta_{\v{k}_s} \zeta_{-\v{k}_s -\v{q}_l} \rangle'$ in the soft limit $q_l \ll k_s$, working to leading order in the slow-roll parameters. According to \eqref{eq:BCA}, we need to find the $ \O(p^2)$ term\footnote{Note that in general, Taylor expansion in $p$ might be ambiguous, at least at first sight. On one hand, we can impose a constraint $\v{k'} = -\v{k} - \v{q} - \v{p}$ and then write expansion coefficients as functions of $\v{q}$ and $\v{k}$. On the other hand - just as an example - we can define $\v{s} := \v{k} + \frac{1}{2} \v{p}$, impose $\v{k'} = -\v{s} - \v{q} -\frac{1}{2} \v{p}$ and then write expansion coefficients as functions of $\v{q}$ and $\v{s}$. The two results will generically differ at order $p^2$ provided that the lower order Taylor coefficients (namely, $\O(p^0)$ and $\O(p^1)$) do not vanish. Nonetheless, it turns out that in the case under consideration ambiguities can be neglected, as we will show in Appendix \ref{app:CanTrispectrum}.\label{ft}} in $P(p)^{-1} \langle  \zeta_{\v{p}} \zeta_{\v{q}} \zeta_{\v{k}} \zeta_{\v{k'}} \rangle$ and sum over directions of $\v{p}$.

For the theory with the action (\ref{eq:CanAction}), the following diagrams contribute to the scalar trispectrum:
\begin{itemize}
\item the contact interaction \cite{Seery:2006},
\item the scalar-exchange diagram,
\item the graviton-exchange diagram \cite{Seery:2008ax}.
\end{itemize}
The scalar-exchange diagram is subleading in the slow-roll parameters \cite{Seery:2008ax}. In Appendix \ref{app:CanTrispectrum}, we show that the graviton exchange contribution vanishes identically after summing over the directions of $p$. In the same appendix we show that the contact contribution starts at order $p^2$ - hence we avoid the ambiguities described in footnote \ref{ft} - and we compute the relevant coefficient.

The final result for small $K$, in the regime $q_l \ll k_s$, to leading order in slow-roll parameters, is
\begin{equation}
\boxed{
\langle \zeta_{\v{q}_l} \zeta_{\v{k}_s} \zeta_{-\v{k}_s} \rangle'_{K} \cong P(q_l)P(k_s) \left[ (1-n_s)  + \O(q_l/k_s) +  \frac{27}{16}  \epsilon \frac{K}{k_s^2} \left( 14 (\hvec{k}_s \cdot \hvec{q}_l)^2 - 13 \right) \right].}
\label{eq:BispectrumCanonicalInflation}
\end{equation}
The power spectra in the above expression are the \textit{flat} power spectra, i.e. those evaluated in the absence of curvature. The $\O(q_l/k_s)$ terms are fixed by the flat-space soft theorem and are not directly affected by $K$. Note that in canonical single-field inflation, the $\O(K)$ correction to the squeezed bispectrum is strongly suppressed by the slow-roll parameter $\epsilon$.

It is arguably more elegant to express the right hand side in terms of curved universe quantities, so rather than use the flat universe power spectrum, we should use $P_K(k)$ (the curved-universe power spectrum evaluated for the canonical single-field scenario). The slow-roll parameter $\epsilon$ can be neglected relative to the scalar tilt $(1-n_s)$ because current data already imposes the small hierarchy $  \epsilon/(1-n_{s})<1/6 $ (which in turn implies conformal invariance of all correlators \cite{Pajer:2016ieg}), we have $P_K(k) = P(k) \left( 1 - \frac{K}{k^2}\right)$ and
\begin{equation}
\boxed{
\langle \zeta_{\v{q_l}} \zeta_{\v{k}_s} \zeta_{-\v{k}_s} \rangle'_{K} \cong P_K(q_l)P_K(k_s) \left[ (1-n_s)  + \O(q_l/k_s) + (1-n_s)\frac{K}{q_l^2} \right].}
\label{eq:BispectrumCanonicalInflation2}
\end{equation}

 
\subsection{An explanation of the scaling in the squeezed bispectrum}\label{ssec:}

In this section, we give a heuristic derivation of the $c_s$ dependence in the squeezed bispectrum $B(q_l, k_s, k_s')$, $q_l \ll k_s \sim k_s'$. We wish to compare the behaviour of the $\O(q_l^2 / k_s^2)$ terms to that of the leading curvature terms $\O(K / k_s^2)$, mainly to demonstrate how the latter effect is enhanced (relative to the former) by the small speed of sound.

We begin by summarizing the size of the standard, flat-universe corrections of order $q_{l}^{2}/k_{s}^{2} $ in the EFT of inflation \cite{Creminelli:2013cga}:
\begin{align}
\ex{\zeta_{\v{q}_{l}}\zeta_{\v{k}_{s}-\v{q}_{l}/2}\zeta_{\v{k}_{s}+\v{q}_{l}/2}}\sim P(q_{l})P(k_{s})\frac{q^{2}_{l}}{k_{s}^{2}} \frac{c_s^{2}-1}{c_{s}^{2}} \left[ (2 + \frac{1}{2}c_3 + \frac{3}{4} c_s^2) - \frac{5}{4} ( \hvec{q}_l \cdot \hvec{k}_s )^2 \right]+\dots\,.
\label{eq:FlatBispectrumCorr}
\end{align}
We note the $c_s^{-2}$ enhancement for $  c_{s}\to 0 $. This scaling can actually be understood without performing the full calculation by tracking the powers of $  c_{s} $.

After re-introducing the term enforced by Maldacena's consistency relation, we schematically have for $  q\ll k $
\begin{equation}
\ex{\zeta_{\v{q}_{l}}\zeta_{\v{k}_{s}-\v{q}_{l}/2}\zeta_{\v{k}_{s}+\v{q}_{l}/2}} \sim P_{u}P_{l} \left[  (1-n_s) +  \O(1) c_s^{-2} \frac{q_{u}^{2}}{q_{l}^{2}} \right].
\end{equation}

Let us now discuss the $\O(K)$ correction to the squeezed bispectrum, which might arise from the \textit{double} soft limit of the trispectrum according to formula (\ref{eq:BCA}). We are only concerned with the double soft limit at $0^{th}$, $1^{st}$ and $2^{nd}$ order in the ultra-long momentum $q_u$. We expect that for a small speed of sound $c_s$, the $\O(q_u^2/k_s^2)$ contribution to the double-squeezed trispectrum will be enhanced by some negative powers of $c_s$.

Here we give a transparent argument for the scaling of the $\O(q_u^2/k_s^2)$ term and reproduce that scaling by tracing the origin of the term to concrete diagrams. At tree-level, the trispectrum receives three contributions: scalar exchange, graviton exchange and contact interaction. Let us discuss them in turn.

\begin{figure}
\begin{center}
\includegraphics[scale=0.12]{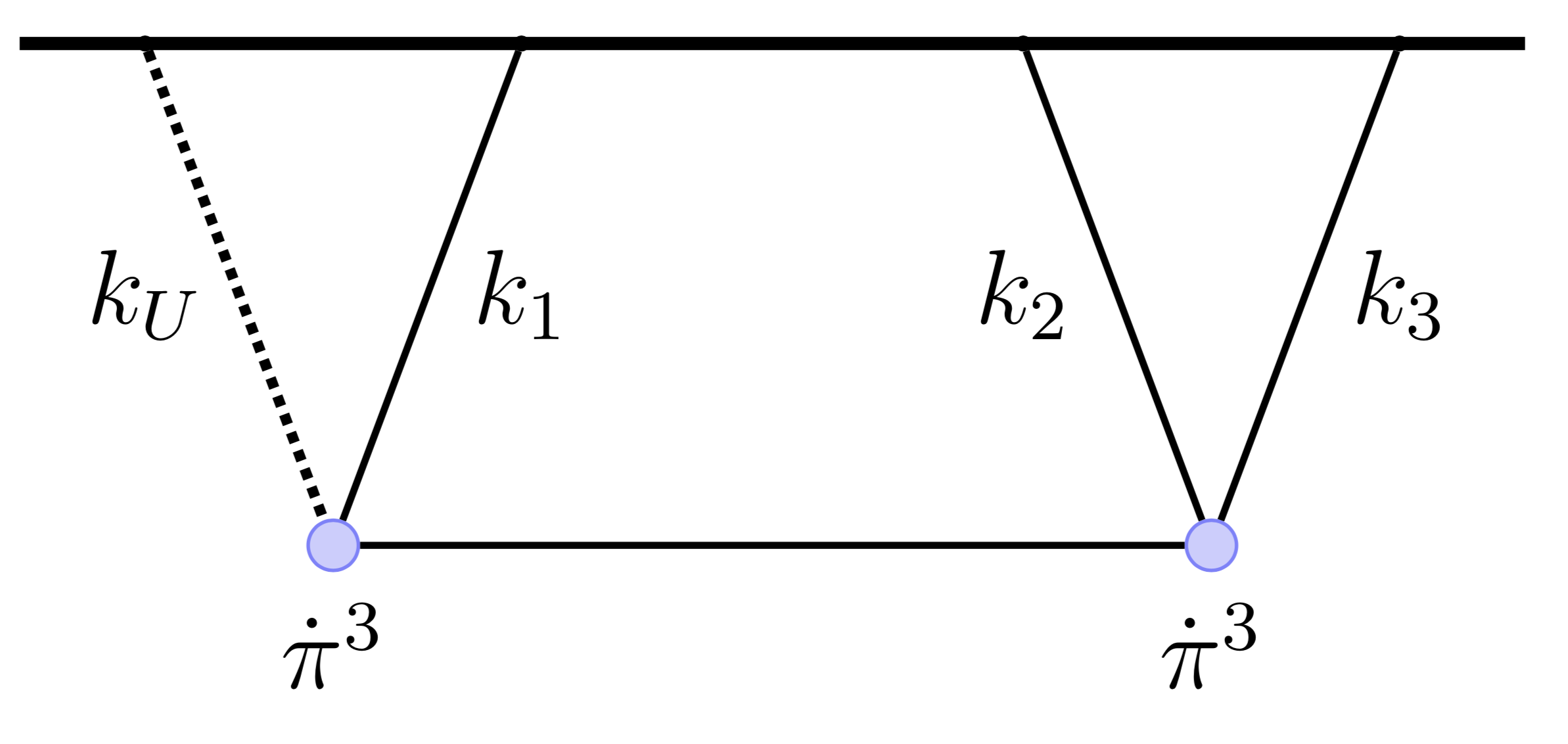}
\includegraphics[scale=0.129]{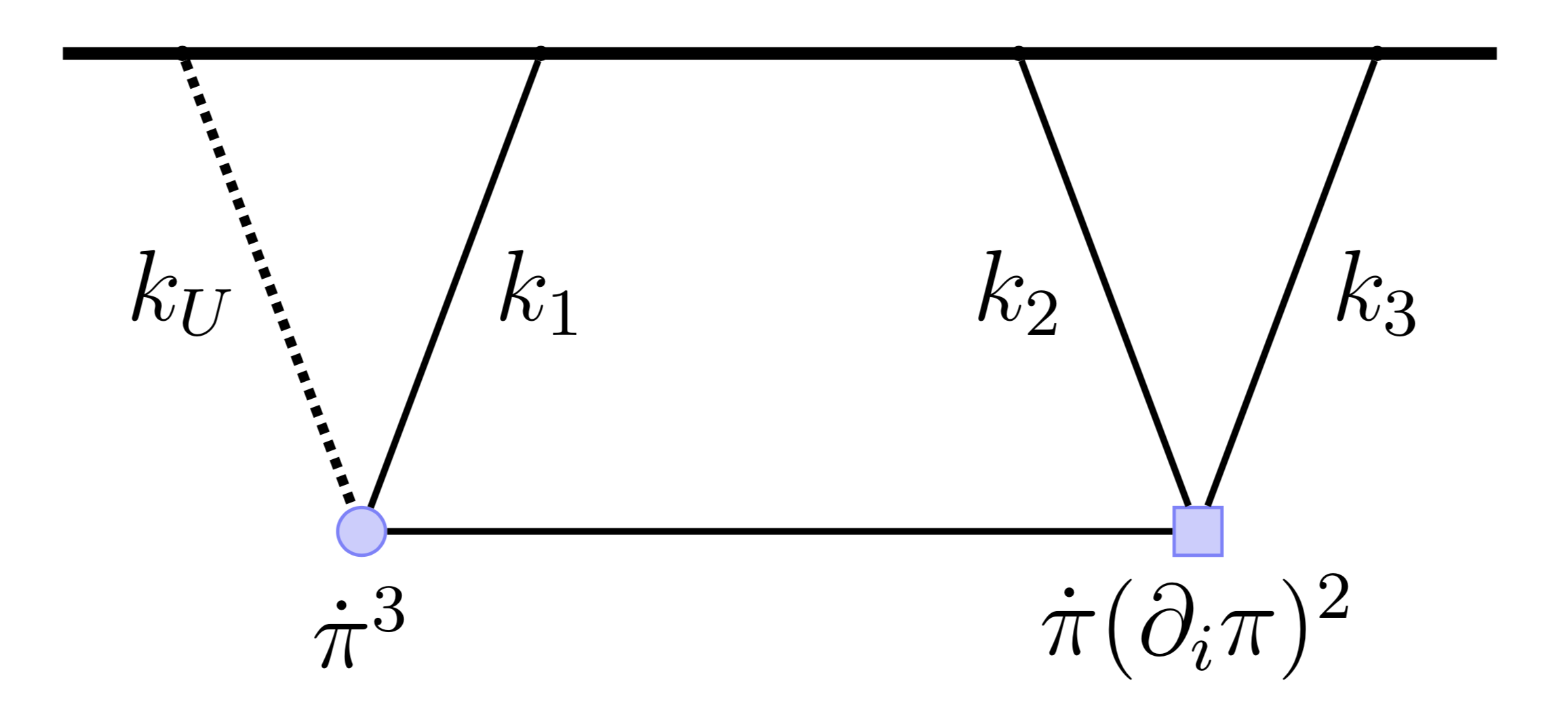}
\includegraphics[scale=0.134]{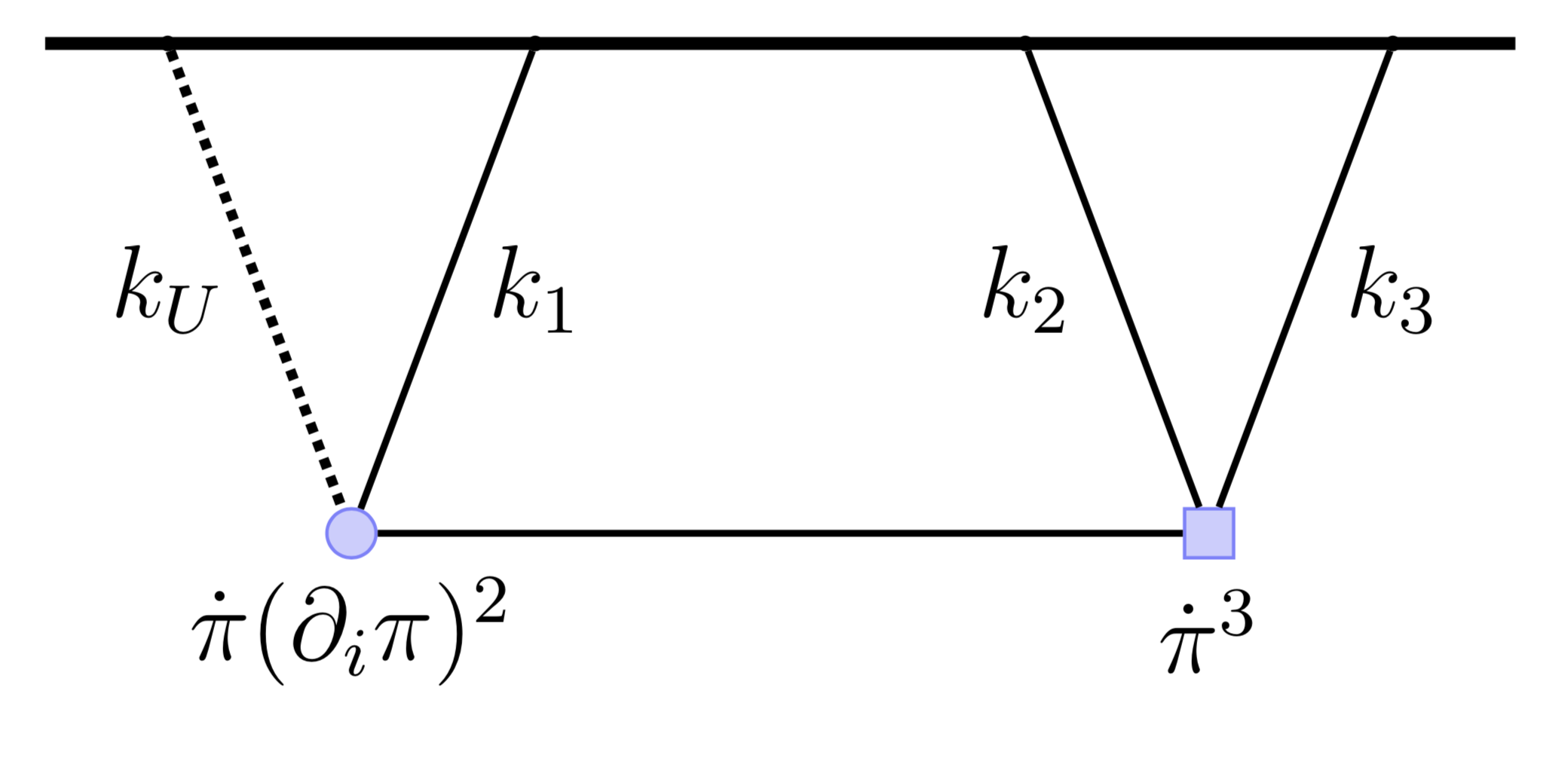}
\includegraphics[scale=0.12]{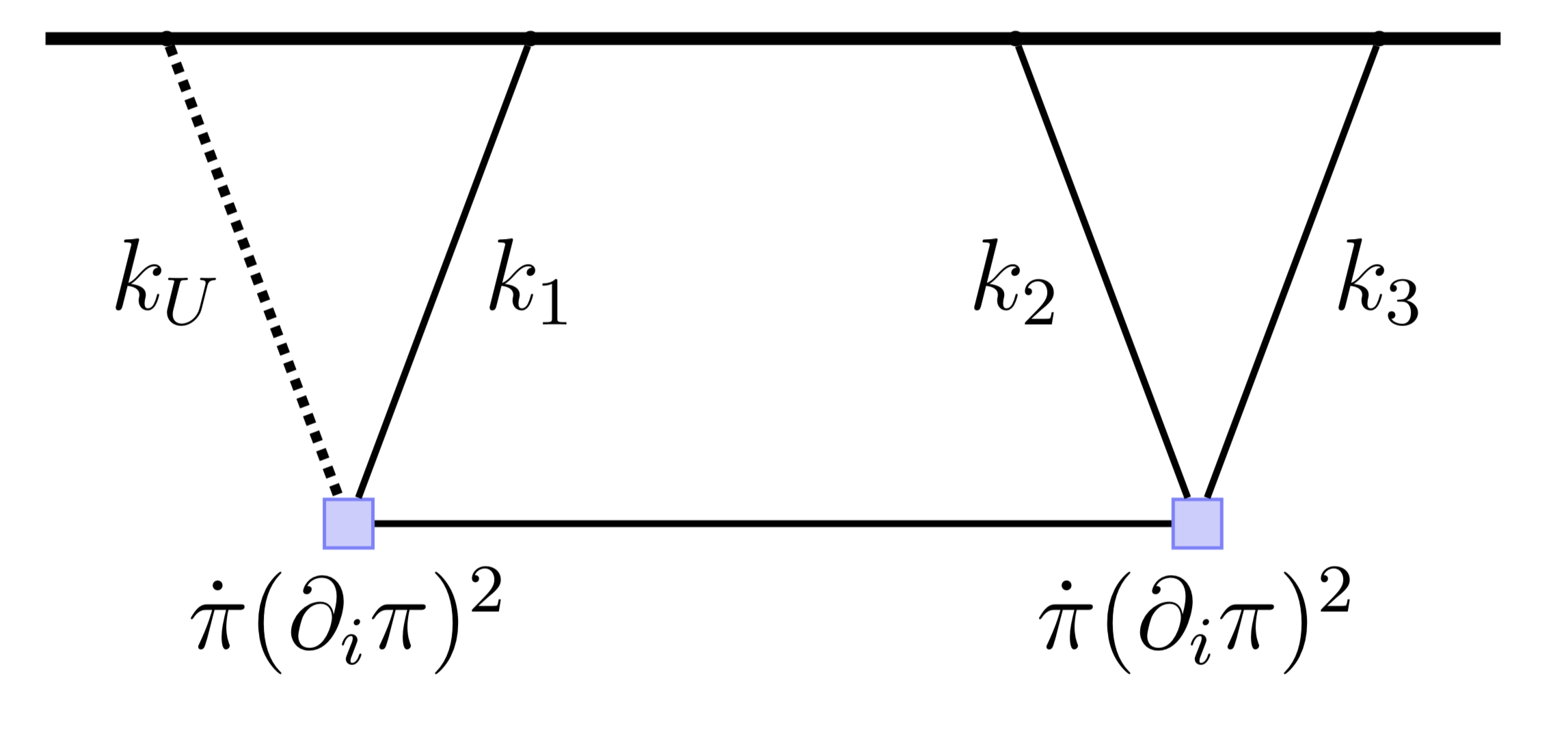}
\includegraphics[scale=0.15]{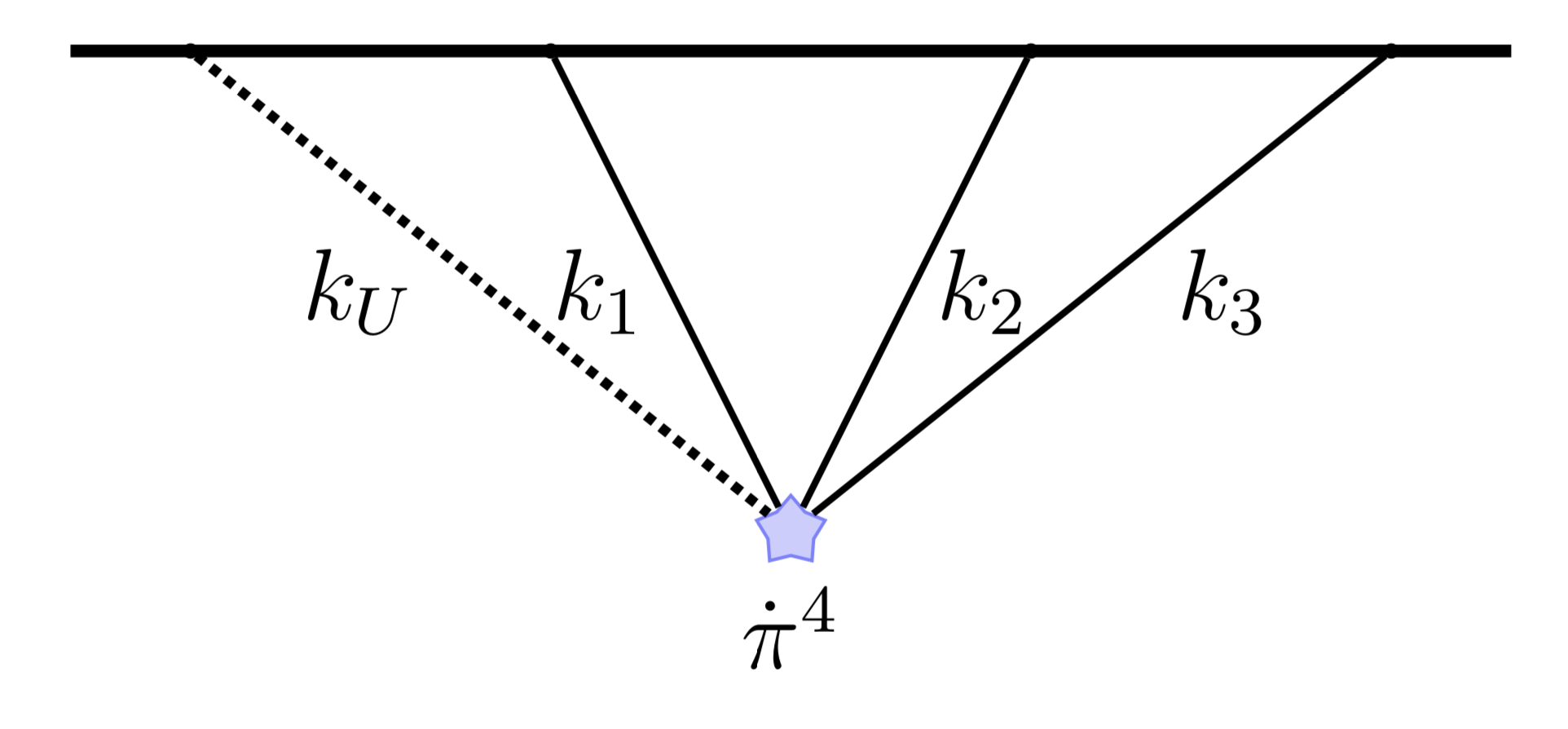}
\end{center}
\caption{The scalar exchange and contact diagrams for the four-point function in the EFT of inflation. $k_U$ corresponds to the ultra-long mode that mimics the effect of spatial curvature.  \label{feynmans}}
\end{figure}

\paragraph{Scalar exchange} This diagram can have poles in the total ``energy'' $ k_{t}=\sum_{a=1}^{4}k_{a} $ and in the momentum $  k_{I} $ of the exchanged scalar. The largest contribution comes from exchanging the softest possible particles, corresponding to the s-channel with $  q_{u} $, $  q_{l} $ coming in, merging into a scalar propagator $  k_{I}\sim q_{l} $ and coming out as $  \vec k_{s} \sim -\vec k_{s}' $. There are two cubic interactions within the diagram, and for each one we get
\begin{align}
q_{u}+q_{l}\to k_{I}\sim q_{l} &\then P_{u}P_{l} \left[  (1-n_{s})+\frac{1-c_{s}^{2}}{c_{s}^{2}} \frac{q_{u}^{2}}{q_{l}^{2}} \right]\,, \\
k_{s}+k_{s}\to k_{I}\sim q_{l} &\then P_{s}P_{l} \left[  (1-n_{s})+\frac{1-c_{s}^{2}}{c_{s}^{2}} \frac{q_{l}^{2}}{k_{s}^{2}} \right] \,.
\end{align}
The resulting trispectrum is (notice that $  P_{l} $ comes only once as it is in the internal ``propagator'' that connects the two vertices)
\begin{equation}
T\sim  P_{u}P_{l} P_{s} \left[  (1-n_{s})^{2}+\frac{\left( 1-c_{s}^{2} \right) (1-n_{s})}{c_{s}^{2}} \left( \frac{q_{u}^{2}}{q_{l}^{2}} +\frac{q_{l}^{2}} {k_{s}^{2}} \right)+\frac{1-c_{s}^{2}}{c_{s}^{4}}\frac{ q_{u}^{2}}{k_{s}^{2}} \right] \,.
\end{equation}

\paragraph{Graviton exchange} While in general graviton exchange does contribute to the squeezed bispectrum, we will show that the contribution always vanishes if we sum over the directions of $\v{k}_{u}$, as we need to do if we want to interpret the ultra-long mode as a spatial curvature. 

We have already seen that the graviton exchange contribution to the trispectrum vanishes after summing over directions of the ultra-long mode in the case of canonical single-field inflation, but this fact holds more generally. Consider the cubic vertex in the graviton-exchange diagram. This vertex is of the form $\pi^2 \gamma$, possibly with some derivatives. Now, $\pi$ is a scalar, but $\gamma \equiv \gamma_{ij}$ is a transverse traceless tensor. We need to contract the $i, j$ indices with derivatives $\partial_i, \partial_j$, and since $\gamma_{ij}$ is transverse, the only lowest order self-consistent operator, up to total derivatives, is $(\partial_i \pi )( \partial_j \pi ) \gamma_{ij}$. (Higher-order operators can be constructed by acting with additional time derivatives or pairs of spatial derivatives.)

The $\gamma_{ij}$ field operator gives rise to a polarization tensor $\epsilon^s_{ij}(\v{k})$ in the correlator and $\partial_i$ gives rise to momentum $k_i$. The vertex factor of our lowest-order operator when one of the $\pi$ legs is the ultra-long mode will be thus proportional to
\begin{equation}
q_i k_j \epsilon^s_{ij}(\v{q} + \v{k}) = - q_i q_j \epsilon^s_{ij}(\v{q} + \v{k})\,,
\end{equation}
which can be shown to always vanish after summing over three orthogonal directions of $\v{q}$.

\paragraph{Contact interaction} The contact interaction has only poles in $  k_{t} $, which goes as  $k_{t}\sim k_{s}+k_{s}' $ in the double squeezed limit. One cannot have any $  1/q_{l} $ enhancement, i.e. the contact interaction can never give any $\O( K/q_{l}^{2} )$ contribution to the squeezed bispectrum.

This interaction might be universal or model dependent, as in the EFT of inflation. The universal part must obey a soft theorem
\begin{align}
T^{\text{grav}} \sim P_{u}P_{l}P_{s} (1-n_{s})\left[ 1+\frac{q_{u}^{2}}{k_{s}^{2}} \right]\dots
\end{align}
where $  (1-n_{s}) $ is a proxy for slow-roll suppressed terms that arise when performing a scaling transformation on the bispectrum. Note the absence of $  q_{u}^{2}/q_{l}^{2} $ terms, due to the fact that there can be no poles in the long momentum $q_l$.

The model-dependent part on the other hand, can contribute to the squeezed limit
\begin{align}\label{eft}
T^{EFT} \sim C P_{u}P_{l}P_{s} \frac{q_{u}^{2}}{k_{s}^{2}}\,,
\end{align}
with some overall amplitude $ C $ that can be large ($ C\gg (1-n_{s}) $). In the squeezed bispectrum, this leads to
\begin{align}
B\sim C P_{l}P_{s} \frac{q_{u}^{2}\zeta_{u}}{k_{s}^{2}} \sim C P_{l}P_{s} \frac{K}{k_{s}^{2}}\,.
\end{align}
While this is the general expectation, single field inflation is an exception. In this case, as we mentioned before, it is only the operator $  \dot \pi^{4} $ that can give a large trispectrum, and the result is 
\begin{align}
T\cong  \frac{1}{q_{1}q_{2}q_{3}q_{4}k_{t}^{5}}\,,
\end{align}
which does not have the form in \eqref{eft}, but rather in the double squeezed limit it behaves as
\begin{align}
T^{EFT}\sim C P_{u}P_{l}P_{s} \frac{q_{u}^{2}}{k_{s}^{2}}\frac{q_{l}^{2}}{k_{s}^{2}}\,.
\end{align}

We quote a more explicit result \cite{Chen:2009se} in Appendix \ref{contactint}. The effect on the bispectrum is then to give a very small correction in the soft limit:
\begin{align}
B \sim C P_{l}P_{s} \frac{K}{k_{s}^{2}} \frac{q_{l}^{2}}{k_{s}^{2}}\, ,
\end{align}
which does not violate Maldacena's consistency relation.


\section{Observational constraints on curvature}\label{sec:ObsSignatures}


In this section, we discuss the possibility of constraining or detecting spatial curvature through measurements of the $\O(K/c_{s}^{2})$ corrections to the primordial power spectrum and bispectrum. In the case of slow-roll, canonical single-field inflation, the corrections are $\O(K/k_s^2)$ and suppressed by a factor of $(1-n_s)$; it is very unlikely that we could detect this signal in the conceivable future. However, in the EFT of inflation with a small speed of sound, $  c_{s}\ll1 $, curvature effects can potentially be observable. We will estimate how large curvature corrections can be given the current separate constraints on curvature and primordial non-Gaussianity. 

Our treatment of curvature corrections is valid at linear order in $  K $. When curvature corrections become large, we can no longer neglect higher-order effects in $  K/c_{s}^{2} $. We will find that current bounds allow for a large magnitude of the $\O(K)$ corrections. This means that the curvature effects \textit{might} be significant and their comparison with observation could provide further constraints on curvature and the speed of sound. On the other hand, we will find that for the power spectrum one needs to go beyond the analysis in this paper and perform a non-perturbative calculation in $  K/c_{s}^{2} $.


\subsection{Constraints from the power spectrum} \label{5p1}

\begin{figure}
\centering
\begin{minipage}[t]{.5\textwidth}
  \centering
  \captionsetup{width=.8\linewidth}
  \includegraphics[width=7 cm]{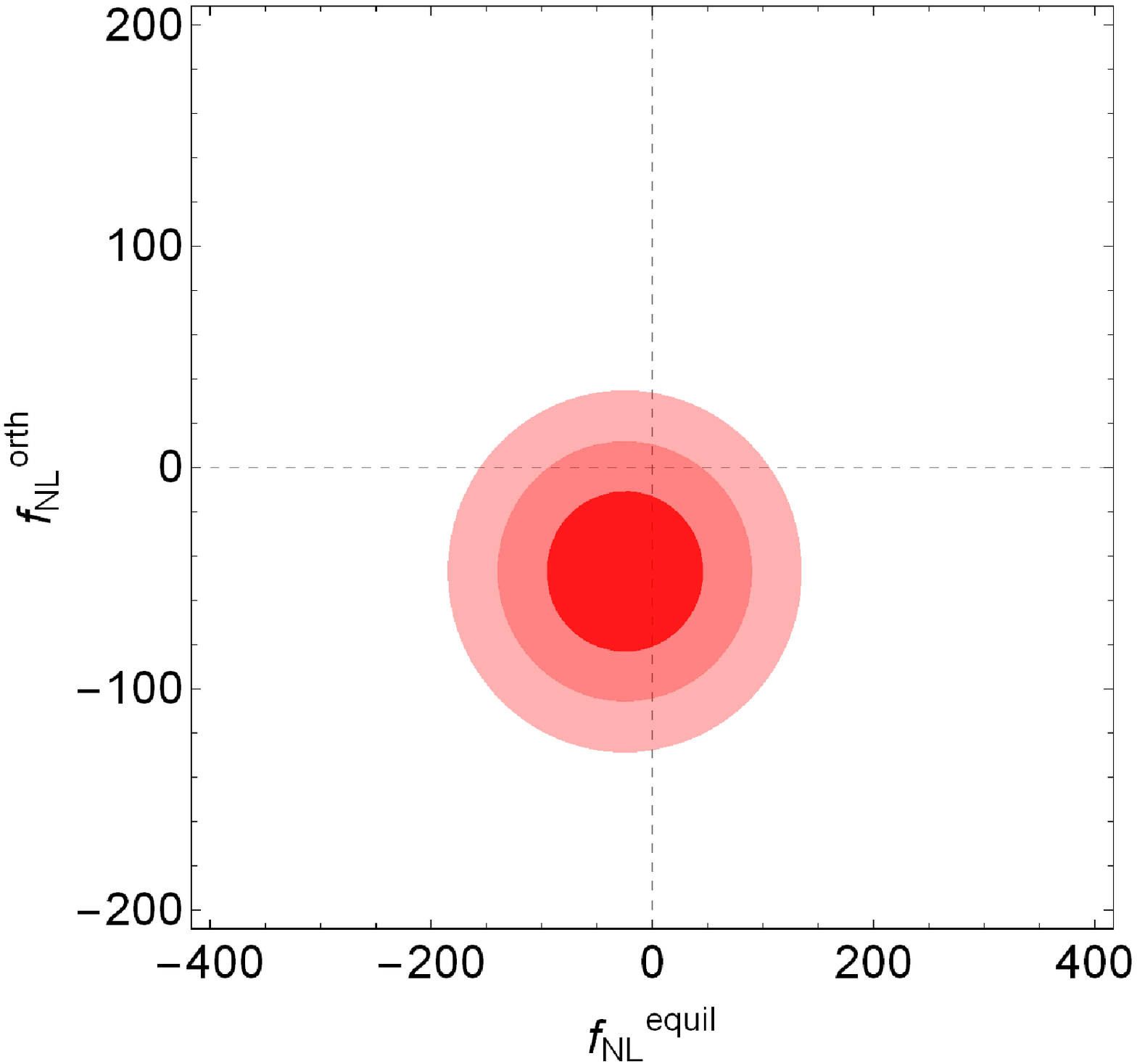}
  \caption{$68\%, 95\%$, and $99.7\%$ confidence regions in the parameter space $(f^{equil}_{NL}, f^{orth}_{NL})$.}
  \label{fig:FNLplot}
\end{minipage}%
\begin{minipage}[t]{.5\textwidth}
  \centering
  \captionsetup{width=.8\linewidth}
  \includegraphics[width=7.3 cm]{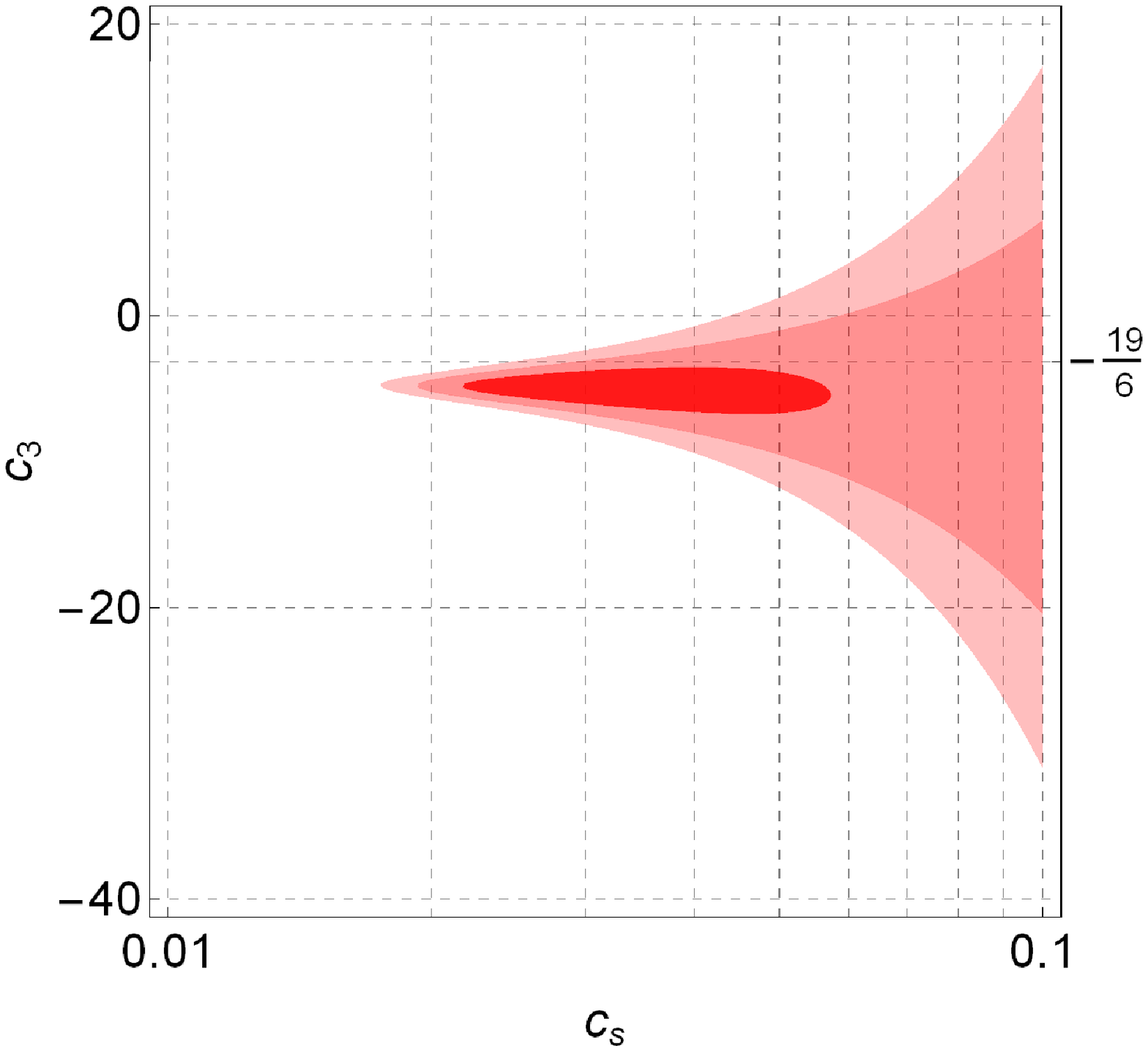} 
  \caption{$68\%, 95\%$, and $99.7\%$ confidence regions in the single-field inflation parameter space $(c_s, c_3)$, obtained from Figure \ref{fig:FNLplot} via the change of variables in Eq. (\ref{eq:fNLequil}) - (\ref{eq:fNLorth}), while allowing $c_s > 1$.}
  \label{fig:Cplots}
\end{minipage}
\end{figure}

As we saw in Sec.~\ref{ssec:SPS}, the curvature corrections to the power spectrum are enhanced in the presence of large non-Gaussianity. Before putting bounds on these corrections, we derive here the current bounds coming from the CMB bispectrum on non-Gaussian parameters and the related coefficients in the EFT of inflation. The EFT of inflation (in a flat universe) predicts the following equilateral and orthogonal non-Gaussianities in the bispectrum \cite{Ade:2015ava}:
\begin{eqnarray} \label{eq:fNLequil}
f^{equil}_{NL} & = & \left(c_s^{-2} - 1 \right) \left[ -0.275 - 0.0780  \left( c_s^2 + 2/3 c_3 \right) \right]   , \\
f^{orth}_{NL} & = & \left(c_s^{-2} - 1 \right) \left[ 0.0159 + 0.0167 \left( c_s^2 + 2/3 c_3 \right) \right]  .  
 \label{eq:fNLorth}
\end{eqnarray}
The Planck collaboration \cite{Ade:2015ava} found (for T+E channels, with lensing not substracted):
\begin{align}
f^{equil}_{NL} &=  - 25 \pm 47, & f^{orth}_{NL} &= - 47 \pm 24  .
\end{align}
We use this observational data to find and plot the $68\%, 95\%$ and $99.7\%$ confidence regions for $f_{NL}^{equil}$ and $f_{NL}^{orth}$ under a simplifying assumption that the $f_{NL}$ covariance matrix is diagonal, which is a good approximation (Figure \ref{fig:FNLplot}). Next, we use (\ref{eq:fNLequil}) - (\ref{eq:fNLorth}) to map the confidence regions to the $\v{c} \equiv (c_s, c_3)$ parameter space (Figure \ref{fig:Cplots}), by noting that for any region $ A  $ in the parameter space, $\mathbb{P}(\v{c}(\v{f}_{NL}) \in \v{c}(A)) = \mathbb{P}(\v{f}_{NL} \in A)$ (we allow $c_s > 1$ for sake of simplicity; in fact, most of the relevant confidence regions do lie within $c_s < 1$). We see that the bispectrum likelihood peaks in the region $0.02 < c_s < 0.1$, $c_3 \sim \O(1)$.

Then we would like to constrain the curvature-induced modification of the power spectrum, given in \eqref{eq:PScurved}, based on the measurements of the CMB temperature angular power spectrum. Let us begin by estimating the 
signal-to-noise ratio in $C_l^{\text{TT}}$'s, 
\be
(S/N)_2\equiv \sqrt{\sum_l\, \dfrac{(C_l^{\text{th}}-C_l^{\text{fid}})^2}{(\Delta C_l)^2}}\,,
\ee
where $C_l^\text{th}$ and $C_l^\text{fid}$ are the temperature angular power spectrum derived from \eqref{eq:PScurved} for $K\neq 0$ and $K=0$, respectively, and $\Delta C_l$ is the cosmic variance in $C_l$. 

\begin{figure}
\includegraphics[width = 9.6 cm]{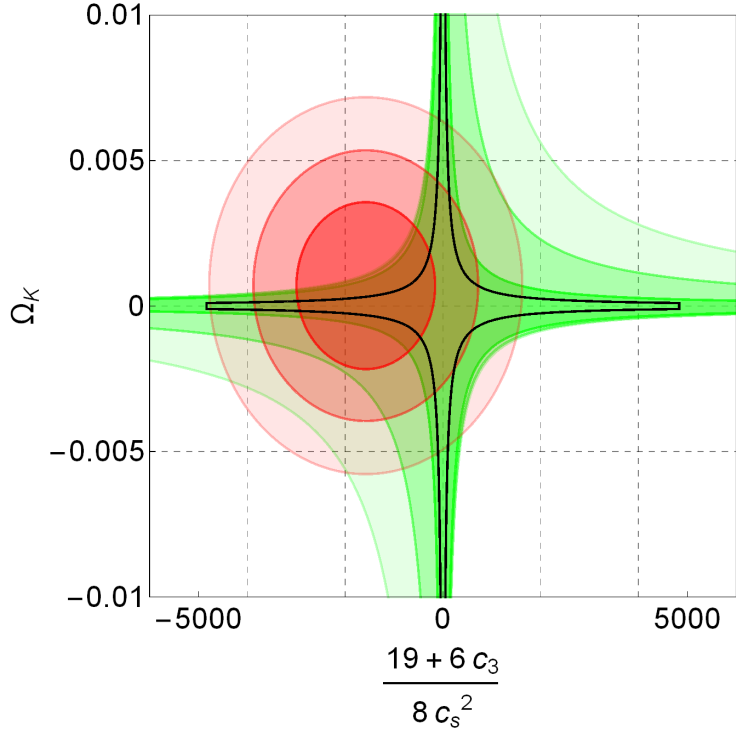}
\centering
\caption{\label{fig:powervsPlanck} The red, circular areas represent the the $68\%$, $95 \%$ and $99.7 \%$ probability regions of the joint PDF of $\Omega_K$ and the non-gaussianity parameter $(19+ 6 c_3)/(8 c_s^2)$, derived from the Planck 2018 data. The green, hyperbolic areas show the $68\%$, $95 \%$ and $99.7 \%$ probability regions constrained by the measurements of $C_{l=2,3,4}$. Finally, the black contour in the centre of the plot shows the approximate region of validity of the linear theory used in this work.}
\end{figure}

As could be anticipated from the scaling of \eqref{eq:PScurved}, $(S/N)_2$ is dominated by the low-$\ell$ multipoles, where we can use the Sachs-Wolfe transfer function to estimate $C_l^\text{th}$.\footnote{Since the dipole ($\ell=1$) is degenerate with the Earth peculiar motion, we discard it. We also neglect slow-roll suppressed terms in our estimation of the signal-to-noise ratio.}  We find
\be
(S/N)_2\sim 3.2\times \Omega_K \, \left( \dfrac{19+6c_s}{8c_s} \right)\, \sqrt{\sum_{\ell>1}\dfrac{4l+2}{9(l^2+l-2)^2}}\sim \Omega_K \,\left( \dfrac{19+6c_s}{8c_s} \right)\,.
\ee
Since more than $ 90\%  $ of signal-to-noise ratio is contained in $l=2,3\,\,\text{and}\,\,4$, we use the latest measurements\footnote{Notice that we use the flat universe prediction for the transfer functions used to compute $  C_{l} $ as opposed to the transfer functions in a curved universe. The difference is only of order $  \Omega_{K} $, much smaller than the $  \Omega_{K}/c_{s}^{2} $ effect that we are after here.} of $C_{2,3,4}$ \cite{Planck-Legacy} and find
\be\label{combo}
\dfrac{ \Omega_K (19+6c_3)}{8c_s^2}=-0.78\,\,^{+1.9}_{-0.6}\,.
\ee 
In Figure \ref{fig:powervsPlanck} we compare this result (green shaded regions) with Planck's latest bounds on $\Omega_K$ and non-Gaussianities\footnote{For $c_s \ll 1$, the linear combination $\alpha f_{NL}^{eq} + \beta f_{NL}^{orth}$ is an unbiased estimator for $z = \frac{3.2(19+ 6 c_3)}{8 c_s^2}$ (for $\alpha = -20.78, \beta = 118.5$).} (red shaded region, where we combined the two independent constraints). Naively, this plot seems to show that within the red confidence region allowed by Planck, there are regions that are excluded by the power spectrum constraints on curvature corrections. However, we should notice that the curvature corrections to the power spectrum in \eqref{eq:PScurved} are derived only at linear order in $  K $ and so they should be trusted as long as 
\be
\label{regime}
\frac{|\Omega_K (19+ 6 c_3)|}{8 c_s^2}\ll 1\,.
\ee
To guide the eye, in Figure \ref{fig:powervsPlanck} we plot a black line where this parameter takes the value $ 1/2$. From the plot it is clear that the validity of our theoretical calculation is slightly more constraining than the CMB temperature power spectrum. In other words, current power spectrum data allows for a curvature correction that goes beyond the linear regime. In order to improve upon Planck's limits on the combination of parameters in \eqref{combo}, one would need to compute the power spectrum to higher order in $K/c_s^2$. This is beyond the scope of this work but it is certainly interesting for future research.

\subsection{Forecast of constraints from the bispectrum}

In this subsection, we revisit the results of Section \ref{sec:SqueezedLimits} and estimate the magnitude of the dominant $\O(K)$ corrections to the bispectrum given the current constraints on $c_s$ and $c_3$.

\subsubsection{Squeezed limit}
Recall the leading-order behaviour of the bispectrum expressed in terms of flat Fourier modes, \eqref{final2}, in the regime $q_l \ll k_s$, $ c_s \ll 1$ and to linear order in $  K $:
\begin{align}
\frac{B(q_l, |\v{k}_s - \frac{1}{2}\v{q}_l |, |\v{k}_s + \frac{1}{2}\v{q}_l |)}{P_{\zeta}(q_l)  P_{\zeta}(k_s)} &\sim (1-n_s) +   \frac{c_s^{2}-1}{c_{s}^{2}}  \left[ (2 + \frac{1}{2}c_3 + \frac{3}{4} c_s^2) - \frac{5}{4} ( \hvec{q}_l \cdot \hvec{k}_s )^2 \right]  \frac{q_l^2}{k_s^2}  \nonumber \\
& \quad +   \left[ B_{K1} +  B_{K2} ( \hvec{q}_l \cdot \hvec{k}_s )^2 \right] \frac{K}{k_s^2} \,,
\end{align}
where
\begin{eqnarray} \label{eq:BK1}
 B_{K1} & \equiv & \frac{3}{16} \left(6  c_3^2 + 43 c_3 + 76 \right) c_s^{-4} = \frac{3}{16} \left(6  c_3 + 19 \right) \left( c_3 + 4 \right) c_s^{-4}, \\
 B_{K2}  & \equiv & - \frac{15}{32} \left(6  c_3 + 19 \right) c_s^{-4}. 
 \label{eq:BK2}
\end{eqnarray}
to leading order in slow-roll coefficients and in $c_s$. The presence of an overall factor of $(6 c_3 + 19)$ in each of the terms is expected, since it arises from the left-hand vertices (those connected to the ultralong mode) in Figure \ref{feynmans}. In the above formula, we neglected contributions due to the contact interactions, originating from terms of the form $\dot \pi^4$ in the EFT action, because they contribute at subleading order, namely $\O(K q_l^2 / k_s^4)$.

The magnitude of the dominant $\O(K)$ correction depends only on the values of the EFT coefficients $c_s$ and $c_3$, which have been partially constrained (see Figure \ref{fig:Cplots} and \ref{fig:powervsPlanck}). We would like to answer two questions:
\begin{itemize}
\item Are the coefficients $B_{K1}$ and $B_{K2}$ sufficiently large for the associated $\O(K)$ effect to have a significant signal-to-noise ratio?
\item Are the coefficients $B_{K1}$ and $B_{K2}$ sufficiently large for the curvature term ($\O(K/k_s^2)$) to be at least comparable to the flat space correction ($\O(q_l^2/k_s^2)$)?
\end{itemize}
To answer the first question, we consider the ratios between the curvature signal and the flat space $\O (q_l^2 / k_s^2)$ contribution to the squeezed limit:
\begin{eqnarray}
R_1 & := & \frac{K}{q_l^2}  \frac{B_{K1} } { \left( 1 - c_s^{-2} \right) \left( 2 + \frac{1}{2}c_3 + \frac{3}{4} c_s^2 \right) } ,  \\
R_2 & := & -\frac{K}{q_l^2}  \frac{B_{K2} } { \frac{5}{4} \left( 1 - c_s^{-2} \right) } .
\end{eqnarray} 
In fact, to leading order in $c_s^{-1}$ we have
\begin{equation}
R_1 = R_2 = \frac{3}{8} \frac{6  c_3 + 19}{c_s^2} \frac{K}{q_l^2} ,
\end{equation} 
so the ratios are identical to the power spectrum effect. We conclude that, within the validity of linear-order treatment of curvature, the curvature corrections cannot be larger than the flat-space corrections ($ \O(q_{l}^{2}/k_{s}^{2})$) to the squeezed limit. It should also be noted that these two corrections have different scaling  and so they are in principle distinguishable.

\begin{figure}
\centering
\includegraphics[width = 0.6 \textwidth]{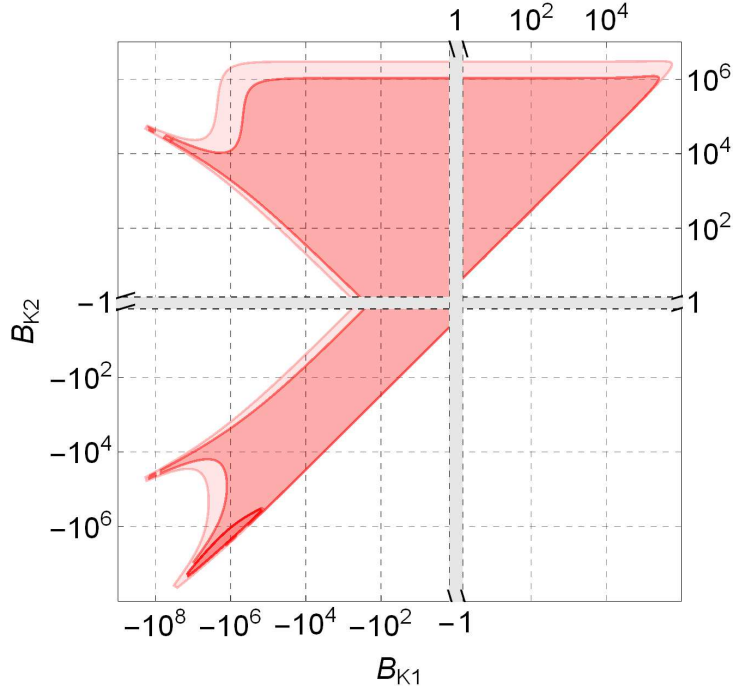}
\captionsetup{width=.9\linewidth}
\caption{$68\%, 95\%$, and $99.7\%$ confidence regions for $(B_{K1}, B_{K2})$, obtained from Figure \ref{fig:Cplots} via change of variables (log scale). Both axes have been snipped so as to show the entire region of interest in a single plot.\label{fig:BK}}
\end{figure}
To answer the second question, we map the constraints (confidence regions) for $c_s$ and $c_3$ obtained from Planck bispectrum data to the constraints (confidence regions) for $B_{K1}$ and $B_{K2}$.\footnote{It is convenient that the map is one-to-one.} Figure \ref{fig:BK} shows the approximate $68\%, 95\%$ and $99.7\%$ confidence regions for $(B_{K1}, B_{K2})$, obtained by mapping the confidence region from the $(c_s, c_3)$ parameter space. We see that the magnitude of the coefficients is bounded, and an order-of-magnitude estimate is $|B_{K1,2}| \lesssim 10^6 \sim 10^7$.
We can then give an upper bound to the signal-to-noise ratio of the squeezed component of the $\O(K)$ correction to the bispectrum. For the estimate, we will take the maximal value consistent with the constraints, $B_{K1,2} = 10^7$, and $|\Omega_K| = 10^{-3}$. We obtain
\begin{align}
\left( \frac{S}{N} \right)_{\text{squeezed}} &\simeq \sqrt{\int_{H_{0}}^{\infty} dq_{l} \,q_{l}\,\int_{k_{s}^{\text{min}}}^{\infty} dk_{s} \int_{-1}^{1} d\cos\theta  \,\frac{B_{K1,K2}^{2} P_{\zeta}^2(q_{l})P_{\zeta}^2(k_{s}) }{P_{\zeta}(q_{l})P_{\zeta}^{2}(k_{s})} \left( \frac{|K|}{k_{s}^2} \right)^{2} } \\
& \sim \, |B_{K1,K2} \Omega_{K,0}| \Delta_{\zeta} \left( \frac{H_{0}}{k_{s}^{\text{min}}} \right)^{3/2} \ll \,10^7  \cdot 10^{-3} \cdot 10^{-5}  \sim 0.1,
\end{align}
where $ k_{s}^{\text{min}}$ is the smallest $  k_{s} $ we want to allow in the squeezed limit $  k_{s}^{\text{min}} \gg q_{l}\geq H_{0}$. Notice that we didn't impose the requirement that the linear term in $  K $ is smaller than the zeroth order term. We conclude that, accounting for current bounds, the curvature corrections to the squeezed bispectrum are too small to be detected, even if we were able to extrapolate the leading-order corrections beyond its regime of validity.

\begin{figure}
\includegraphics[width=11 cm]{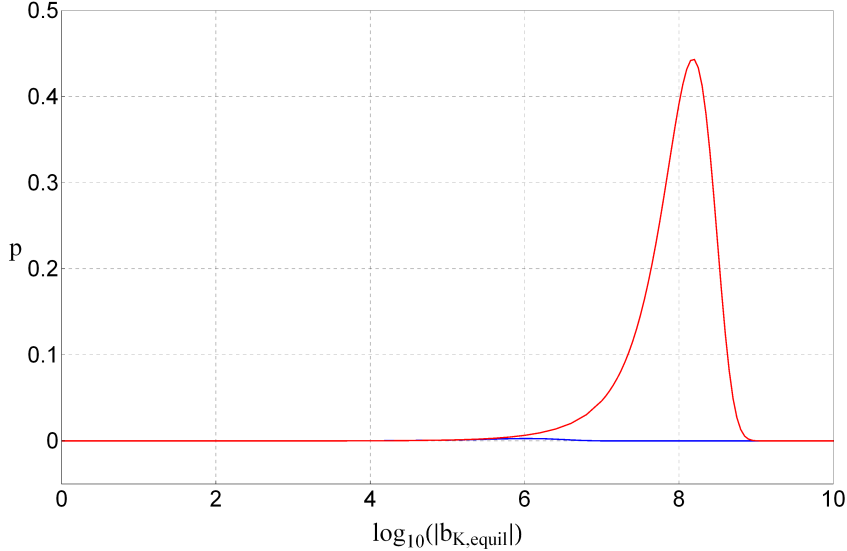}
\centering
\caption{The posterior probability distribution function (unnormalized) of the $\text{sign}(b_{K,\text{equil}})\log_{10}(|b_{K,\text{equil}}|)$ parameter, derived from the Planck bispectrum data, assuming a flat prior on $(f_{NL}^{\text{equil}}, f_{NL}^{orth})$. Blue line shows the pdf for $b_{K,\text{equil}} > 0$ while the red line shows the pdf for $b_{K,\text{equil}} < 0$. \label{fig:BKequil}}
\end{figure}

\subsubsection{Equilateral configuration}
Linear curvature correction to the bispectrum is suppressed by $k_s^{-2}$ in the squeezed limit, so we may hope to obtain a larger signature if we consider an equilateral configuration of the momenta. We use the same background-curvature argument as in Section \ref{sec:SqueezedLimits}. We need to consider the trispectrum in the limit where one of the modes becomes very long, while the other three form an equilateral shape, $q_u \ll k_2 = k_3 = k_4 \equiv k$:
\begin{equation}
\langle \zeta_{\v{q}_u} \zeta_{\v{k}_2} \zeta_{\v{k}_3}  \zeta_{\v{k}_4} \rangle\, = \langle \zeta_{\v{q}_u} \zeta_{\v{k}_2} \zeta_{\v{k}_3}  \zeta_{\v{k}_4} \rangle\,_{SE} + \langle \zeta_{\v{q}_u} \zeta_{\v{k}_2} \zeta_{\v{k}_3}  \zeta_{\v{k}_4} \rangle\,_{GE} + \langle \zeta_{\v{q}_u} \zeta_{\v{k}_2} \zeta_{\v{k}_3}  \zeta_{\v{k}_4} \rangle\,_{CI}.
\end{equation}
The graviton exchange does not contribute to the bispectrum correction. The scalar exchange and contact interaction diagrams will generically both lead to a significant $\O(K/k^2)$ effect.

The scalar-exchange diagrams give, to leading order in $c_s$:
\begin{align}
B(k,k,k)_{K, SE} = P_{\zeta}(k)^2 \left[ \frac{65}{6} \left( 1+\dfrac{2}{3}\dfrac{c_3}{c_s^2}\right)^2 + \frac{5321}{432} \left( 1+\dfrac{2}{3}\dfrac{c_3}{c_s^2}\right) c_s^{-2}+ \frac{4747}{216}  c_s^{-4} \right] \times \frac{K}{k^2}\,.
\end{align}

The contact interactions give, to leading order in $c_s$:
\begin{eqnarray}
\label{contactterm}
B(k,k,k)_{K, CI} =\dfrac{16 K}{3 k^2} P_{\zeta}(k)^2 \left[\,C_{\dot{\pi}^4}-\dfrac{9}{4}\left(1+\dfrac{2}{3}\dfrac{c_3}{c_s^2}\right)^2\right].
\end{eqnarray}

We first focus on the enhancement that cubic interactions induce by neglecting the $C_{\dot{\pi}^4}$ coefficient. In this case, the curvature correction to the bispectrum in the equilateral configuration becomes
\begin{align}
B(k,k,k)_{K} \approx  P_{\zeta}(k)^2 \left( - 0.519 c_3^2 +  16.42 c_3 + 21.98 \right) c_s^{-4} \times \frac{K}{k^2} .
\end{align}
Let 
\begin{align}
b_{K,\text{equil}} :=  \left( - 0.519 c_3^2 +  16.42 c_3 + 21.98  \right) c_s^{-4}\,.
\end{align}
Assuming a uniform prior for the equilateral and orthogonal non-Gaussianities $f_{NL}^{\text{equil}}$ and $f_{NL}^{orth}$ in the flat-universe approximation and working in the limit $c_s \ll 1$, we find the posterior pdf for $b_{K, \text{equil}}$ by mapping the corresponding pdf for the non-Gaussianities and marginalizing over $c_3$. The posterior pdf for $b_{K, \text{equil}}$ is shown on Figure \ref{fig:BKequil}. We see that $b_{K,\text{equil}} \sim 10^8$ is still allowed by current constraints. 

Let us switch off the cubic terms and consider the contribution of the quartic term $\dot{\pi}^4$ to the equilateral bispectrum,
\be
\dfrac{B_{K,CI}(k,k,k)}{P^2_{\zeta}(k)}\sim \dfrac{16 C_{\dot{\pi}^4}}{3}\dfrac{K}{k^2}\,.
\ee
The Planck observational constraint on the trispectrum \cite{Ade:2015ava} implies the following rough upper bound on the coefficient in front of $K/k_s^2$, 
\be
\dfrac{16}{3}\,|C_{\dot{\pi}^4}|<10^7\,.
\ee
This is slightly weaker than the maximum size allowed for $b_{K,\text{equil}}$ and so it can be neglected for an order of magnitude estimate. Let us now derive a rough upper bound on the signal-to-noise ratio for measuring spatial curvature in the 3D bispectrum of $  \zeta $ as follows:
\begin{align}
\left( \frac{S}{N} \right)_{\text{equil.}} &\simeq \sqrt{\int_{H_{0}}^{\infty} dk \,k^{2\,}  \frac{b_{K,\text{equil}}^{2} P_{\zeta}^4(k) }{P_{\zeta}^{3}(k)} \left( \frac{|K|}{k^2} \right)^{2} } \\
& \sim \, |b_{K,\text{equil}} \Omega_{K,0}| \Delta_{\zeta}\,.
\end{align}
Notice that the integral in $  dk $ is strongly supported on the largest observable scales, $k = H_{0}$, and so it is insensitive to the UV cutoff in $  k $, which we have taken to infinity for simplicity. If we naively used $|b_{K,\text{equil}}| = 10^8$, a proxy for the largest value allowed by the current constraints depicted in Figure \ref{fig:BKequil}, and $ |\Omega_{K,0}|=10^{-3}$, we would get that the signal is barely detectable $  S/N \sim 10^{8-3-5}=1$. From this estimate we conclude that the corrections to the equilateral configurations are unlikely to be detectable even if we extrapolate beyond the validity of the linear-order treatment in $  K $. On the other hand, if we insist that the curvature correction is smaller than the $  K^{0} $ term, then we can at most take $  |b_{K,\text{equil}} \Omega_{K,0}| \sim f_{NL}^{\text{eq,ort}} \sim 10^{2} $. We then find a very small signal-to-noise ratio, $  S/N\sim 10^{-3}$. The intuitive reason why the curvature corrections are so hard to detect is that, although they can be very large for the largest observable scales, the non-scale invariant signal drops very quickly on shorter scales and the signal-to-noise ratio saturates with just a few multipoles. Conversely, the traditional non-Gaussianity such as equilateral and orthogonal shapes are scale invariant and they can be constrained by all $  B_{l_{1},l_{2},l_{3}} $.


\section{Discussion and conclusion}\label{sec:Discussion}

In this work, we have discussed the effects that spatial curvature induces in primordial correlators, at linear order. As explained in Figure \ref{fig1}, these effects are parameterised by $  \Omega_{K}/c_{s}^{2} $ and so could be large if $  c_{s} $ was small during inflation or equivalently if perturbations interacted more strongly. We have shown that in the presence of curvature, the soft limit of the bispectrum acquires model-dependent corrections that deviate from Maldacena's consistency relation in flat space. More generally, we have argued that residual diffeomorphisms are separated from the spectrum of physical perturbations by a gap of order $  |K| $ and so standard soft theorems should be violated at linear order in curvature. We have furthermore studied how large these corrections can be both in the power spectrum and in the bispectrum, also going beyond the squeezed limit. We have found that in the power spectrum, constraints from the CMB are close to but slightly weaker than the validity of our linear order treatment of curvature. For the bispectrum on the other hand, the signal-to-noise for these corrections is always smaller than one, even assuming we had access to a full 3D map of the primordial correlators to the cosmic variance limit.

There are a few avenues for future research. First, as discussed in Sec. \ref{5p1}, we could not harvest the full constraining power of the CMB because our theoretical prediction was limited to linear-order in $  K/c_{s}^{2} $, in which regime the corrections to the temperature angular power spectrum are slightly smaller than the experimental bound. It would therefore be very interesting to compute the power spectrum of primordial perturbations with small $  c_{s} $ in a curved universe to all orders in $  K $ and then compare the prediction again with the lowest CMB multipoles. Such calculation was performed for $  c_{s}=1 $ in \cite{Yamamoto:1995sw,Yamamoto:1996qq,Garriga:1998he}. In that work, the authors also accounted for the initial state dictated by the Euclidean continuation for a bubble nucleation event, that is different from the Bunch-Davies state we have used in this work. The difference is non perturbative in $ K $ and our preliminary analysis shows that it might be safe to neglect this effect, but a more detailed study should be performed. Second, it would also be interesting to use the polarisation of the CMB and analyse $ C_{l}^{EE}$ and $ C_{l}^{TE}$ to improve the constraints on curvature corrections to the power spectrum. Because polarisation is so small on Hubble scales, we do not expect a large improvement from this additional data. In fact, it seems likely that $ C_{l}^{EE} $ will give a negligible improvement because the error bars on the first few $ l$'s are so large, while $ C_{l}^{TE}$ might improve the bounds by a few tens of percent, after the covariance has been taken into account. Third, one might extend our analysis of the CMB power spectrum to the bispectrum, including both temperature and polarisation but one should be aware that our rough estimate for the signal-to-noise ratio in the bispectrum is much smaller than one within the regime of validity of the linear theory; even going beyond this regime, the ratio is at most $\O (1)$ for the largest allowed values of parameters. Finally, in this work we have focussed on single-field inflation because of our interest in discussing the squeezed limit consistency relation. But one could investigate curvature effects in multifield inflation, where there is more room for producing a large signal.

\section*{Acknowledgements} We are thankful to Paolo Creminelli, James Fergusson, Mehrdad Mirbabayi and David Stefanyszyn for useful discussions. S. J. and E. P. have been supported in part by the research program VIDI with Project No. 680-47-535, which is (partly) financed by the Netherlands Organisation for Scientific Research (NWO). J. S. has been supported by a grant from STFC.


\appendix

\section{Perturbations around FLRW} \label{app:FLRW}
\label{FRW}

Here we collect useful formulas on the kinematics of FLRW spacetimes. The non-vanishing Christoffel symbols in the coordinates used in \eqref{flrwm} are
\begin{align}
&\Gamma^0_{ij}= a\dot{a}\tilde{g}_{ij}\,, \\ \nn
&\Gamma^i_{0i}= H\,\delta_{ij}\,,\\ \nn
&\Gamma^k_{ij}= \dfrac{1}{2} K f \left( x^k \delta_{ij}-2 x^{(i}\delta_{j)k}\right)\,.
\end{align}
On spatial sections, the Laplacian operator acting on scalars is
\begin{align}
\nabla^2\,S=\dfrac{1}{f^2}\left(\partial_i\partial_i\,S-\dfrac{1}{2}K\,f\,x^i\partial_i\,S\,\right)\,.
\end{align}
The isometry group of a curved FLRW spacetime is SO(3,1) if it is open, and SO(4) if it is closed. The Killing vectors are
\begin{align}
&T_i=f(2f-1)c_i+\dfrac{K}{2}f^2\,c_jx^j\,x^i\,,\\ \nn
&R_i=f^2\omega_{ij}x^j\,,\qquad \omega_{(ij)}=0\,.
\end{align}
Sometimes the $T_i$'s are called {\it quasi-translations}, i.e. they become ordinary spatial translations in flat space, while the $R_i$'s form rotations.

\subsection{The Scalar-Vector-Tensor Decomposition}
The scalar-vector-tensor decomposition of a tensor field living on a constant curvature manifold with the metric 
\be
ds^2=f^2(K\bfx^2)\,d\bfx \cdot d\bfx\,=f^2(Kr^2)(dr^2+r^2d\Omega^2),
\ee
is an old topic of interest in differential geometry. Here we briefly review the decomposition of a tensor on a sphere or a hyperboloid, along the lines of \cite{Stewart:1990fm}. We would like to determine under what condition the Poisson equation,
\begin{align}
\nabla^2\,\Phi=J(x)\,,
\end{align}
admits a unique solution $  \Phi $ for any given $  J $. If two solutions existed, their difference would solve the Laplace equation, namely
\be
\nabla^2\,\Phi=0\,.
\ee
Multiplying both sides with $\Phi$ and integrating over the whole space, one finds
\be\label{vici}
-\int_{{\cal M}}\,\sqrt{g}\,d^3x\,(\nabla_i\,\Phi)^2+\int_{\partial{\cal M}}\, \sqrt{h}\,d\theta\,d\phi\,\Phi\,n^i\,\nabla^i \Phi=0\,,
\ee
where $g_{ij}$ is the metric of the curved space, $g$ is its determinant, $\partial {\cal M}$ is empty for a sphere and is a 2-sphere for a  hyperboloid. For the latter, $n^i$ is the unit vector normal to the boundary 2-sphere and $h_{ab}$ $(a,b=\theta,\phi)$ is the induced metric on the boundary 2-sphere. Therefore, on a 3-sphere, \eqref{vici} implies that $  \nabla_{i}\Phi=0 $ everywhere, giving as only solution $  \Phi= $ const. Let us move to the hyperboloid. If $\Phi$ decays rapidly enough towards the boundary, namely
\be
f\,\Phi\, \partial_r\,\Phi\to 0\,\quad\text{for}\quad r\to \frac{2}{\sqrt{|K|}}\,,
\ee
then the only solution of the Laplace equation is $  \Phi =$ const, and therefore the solution of the Poisson equation is unique up to a constant.
This in turn implies that the splitting of a vector into a longitudinal and a transverse part, i.e. 
\be
A_i=\nabla_i \phi+A_i^T\,,\qquad \nabla^i\,A_i^T=0\,,
\ee
is unique {\it iff} $\int \sqrt{g}d^3x\,\nabla_i\,A^i$ is finite. 

Rank-2 objects should be dealt with more carefully because the non-commutation of covariant derivatives brings about some complication. Consider the following decomposition
\be
H_{ij}=H_S^{(1)}g_{ij}+\nabla_i\nabla_j\,H_S^{(2)}+2\nabla_{(i}H^V_{j)}+H^T_{ij}\,,
\ee
in which
\be
g^{ij} H^T_{ij}=\nabla^i\,H^T_{ij}=\nabla^i\,H_i^V=0\,.
\ee
It is easy to check that 
\begin{align}
H^i_{\,i}&= 3H_S^{(1)}+\nabla^2\,H_S^{(2)}\,,\\ \nn
\nabla^{-2}\nabla^i\nabla^j\,H_{ij}&=H_S^{(1)}+(\nabla^2+2K)H_S^{(2)}\,. 
\end{align}
Assuming a proper asymptotic decay of $H_{ij}$, $H_S^{(1,2)}$ can be uniquely fixed up to 
\be
\label{ambig}
H_S^{(2)}\to H_S^{(2)}+\chi\,,\quad H_S^{(1)}\to H_S^{(1)}+K\chi\,,
\ee
where $\chi$ is any solution of 
\be
(\nabla^2+3K)\chi=0\,.
\ee 
It remains to solve for $H^V_i$ by taking the divergence of $H_{ij}$,
\be 
2\nabla^i\,\nabla_{(i}H^V_{j)}=\nabla^i\,H_{ij}-\nabla_j\,\Big(\nabla^{-2}\nabla^k\nabla^l\,H_{kl}\Big)\,.
\ee
However, the homogeneous equation, i.e. $\nabla^i\,\nabla_{(i}H^V_{j)}=0$, could admit non-trivial solutions. Multiplying the latter with $H^V_j$ and integrating over the space yields
\be
\int\,\sqrt{g}\,d^3x\, H^{V\,i}\,\nabla^j\nabla_{(i}H^V_{j)}=\int_{\partial M}\sqrt{h}d\theta d\phi\,H^{V\,i}\,\nabla_{(i}H^V_{j)}-\dfrac{1}{2}\int \sqrt{g}d^3x\, (\nabla_{(i}H^V_{j)})^2=0\,. \nonumber
\ee
Therefore, for both a hyperboloid---assuming that $H^V_i$ vanishes quickly enough near the boundary----and a sphere, the boundary term vanishes, and as a result the solutions consist only of the Killing vectors. However, Killing vectors of a hyperboloid are not bounded, hence we must  discard them. In conclusion, the SVT decomposition of $H_{ij}$ is unique up to $\eqref{ambig}$ and separately
\be
H^V_i\to H^V_i+\xi_i\,,\quad \nabla_{(i}\,\xi_{j)}=0\,.
\ee
Notice that for all $\xi_i$'s we have$\nabla^2\xi_i=-2K\,\xi_i$. Thus, as long as one works with eigenfunctions of the $\nabla^2$ operator with eigenvalues unequal to $-3K$ (for scalars) and $-2K$ (for vectors), the SVT decomposition is unique and well defined. 
\subsection{The Spectrum of The Laplacian Operator}
For $K<0$ the eigenfunctions of the Laplacian operator for scalars, defined through
\be
\nabla^2\,Y_{plm}=(1+p^2)K\,Y_{plm}\,,
\ee
are given by (see e.g.\cite{Sasaki:1994yt})
\be
Y_{plm}=\frac{\Gamma(ip+l+1)}{\Gamma(ip+1)}\dfrac{p}{\sqrt{|K|r\,f(Kr^2)}}P^{-l-1/2}_{ip-1/2}\Big(\sqrt{1+K^2r^2f^2(Kr^2)}\Big)\,Y_{lm}(\theta,\phi),\nonumber
\ee
where $P^\mu_\nu(x)$ are Associated Legendre functions of the first kind, and $\Gamma(x)$ is the Euler Gamma function. Only mode functions with $p>0$ ($-\nabla^2>|K|$) are square integrable and in this sense physical. They are also normalized and orthogonal, i.e. 
\be
\label{ortho}
\int\, r^2\,f^3\,drd\Omega\, Y_{plm}Y^*_{p'l'm'}=\delta(p-p')\delta_{mm'}\delta_{ll'}\,.
\ee

For $K>0$ the spectrum of scalar harmonics is (\cite{Ratra:2017ezv})
\be
\nabla^2\,Y_{plm}=-p(p+2)K\,Y_{plm}\,,\quad p=0,1,..\,,\quad \text{and}\,\quad l=0,..,p\,.
\ee
and they are given by 
\be
Y_{plm}=\sqrt{\dfrac{(p+1)\Gamma(p+l+2)}{\Gamma(p-l+1)}}\dfrac{p}{\sqrt{Kr\,f(|K|r^2)}}P^{-l-1/2}_{p-1/2}\Big(\sqrt{1-K^2r^2f^2(Kr^2)}\Big)Y_{lm}(\theta,\phi)\,,\nonumber
\ee
with the same property as \eqref{ortho} except that $\delta_{pp'}$ replaces $\delta (p-p')$. 


\section{Canonical Trispectrum in the soft limit} \label{app:CanTrispectrum}

In this appendix, our goal is to compute the $O(q_{u}^2)$ term in the pure scalar $4$-point function
\begin{equation}
\langle \zeta_{\v{q_{u}}} \zeta_{\v{q_{l}}} \zeta_{\v{k_{s}}} \zeta_{\v{k'_{s}}} \rangle\,.
\end{equation}
 \cite{Seery:2006, Seery:2008ax} give explicit formulas for this $4$-point function derived under the following assumptions:
\begin{itemize}
\item Inflation is described by a single field with Lagrangian $\mathcal{L} = - \frac{1}{2}(\nabla \phi)^2 - V(\phi)$.
\item All interactions are that of the inflaton minimally coupled to Einstein gravity.
\item Terms that are suppressed by higher powers of slow-roll parameters can be neglected, ie. we work to leading order in $\epsilon$ and $\eta$.\footnote{In particular, the tree diagram involving an exchange of a scalar is suppressed by additional power of slow-roll parameters. Thus, at tree level we only need to consider the contact interaction and the graviton-exchange diagram.}
\end{itemize}

The $4$-point function has a contribution due to the contact interaction \cite{Seery:2006} as well as due to graviton exchange \cite{Seery:2008ax}:
\begin{equation}
\langle \zeta_{\v{k}_1} \zeta_{\v{k}_2} \zeta_{\v{k}_3} \zeta_{\v{k}_4} \rangle = \langle \zeta_{\v{k}_1} \zeta_{\v{k}_2} \zeta_{\v{k}_3} \zeta_{\v{k}_4} \rangle_{CI} + \langle \zeta_{\v{k}_1} \zeta_{\v{k}_2} \zeta_{\v{k}_3} \zeta_{\v{k}_4}\rangle_{GE}.
\end{equation}
We are interested in the double squeezed limit, and we will take $\v{q}_u \equiv \v{k}_1$, $\v{q}_u \equiv \v{k}_2$, $\v{k}_s \equiv \v{k}_3$, $\v{k}'_s \equiv \v{k}_4$. We have to compute the term proportional to $q_u^2$.

\subsection{The graviton exchange}
The contribution to the trispectrum from the graviton exchange is given by \\
\begin{align}
\label{eq:GEcrude}
\langle \zeta_{\v{k}_1} \zeta_{\v{k}_2} \zeta_{\v{k}_3} \zeta_{\v{k}_4} \rangle^{GE} = (2 \pi)^3 \delta(\sum\limits_a \v{k}_a) \frac{H_*^6}{\epsilon^2 \prod_a (2 k_a^3)}  k_1^2  \\ 
\nonumber
\times \left[
\frac{k_3^2}{k_{12}^3} [1 - (\hvec{k}_1 \cdot \hvec{k}_{12})^2] [1 - (\hvec{k}_3 \cdot \hvec{k}_{12})^2]  \cos 2 \chi_{12,34} \cdot (\mathcal{I}_{1234} + \mathcal{I}_{3412})
\right. \\ 
\nonumber
\left. +
\frac{k_2^2}{k_{13}^3} [1 - (\hvec{k}_1 \cdot \hvec{k}_{13})^2] [1 - (\hvec{k}_2 \cdot \hvec{k}_{13})^2] \cos 2 \chi_{13,24} \cdot (\mathcal{I}_{1324} + \mathcal{I}_{2413}) \right. \\ 
\nonumber
\left. +
\frac{k_2^2}{k_{14}^3} [1 - (\hvec{k}_1 \cdot \hvec{k}_{14})^2] [1 - (\hvec{k}_2 \cdot \hvec{k}_{14})^2] \cos 2 \chi_{14,23} \cdot (\mathcal{I}_{1423} + \mathcal{I}_{2314})
\right]
\end{align}

where $\mathcal{I}_{abcd} + \mathcal{I}_{cdab}$ are given below, and $\chi_{12,34}$ is the angle between the plane defined by the vectors $\v{k}_1$, $\v{k}_2$ and the plane defined by the vectors $\v{k}_3$ and $\v{k}_4$. Although the above formula may not look \textit{manifestly} invariant under relabelling of the momenta, we verified it to be unaffected by permutations.    \\
\begin{align} 
\mathcal{I}_{1234} + \mathcal{I}_{3412} = \frac{k_1 + k_2}{a_{34}^2} \left[ \frac{1}{2} (a_{34} + k_{12}) (a_{34}^2 - 2 b_{34}) + k_{12}^2 ( k_3 + k_4 ) \right] + (1,2 \leftrightarrow 3,4) \\ \nonumber
+ \frac{k_1 k_2}{k_t} \left[ 
\frac{b_{34}}{a_{34}} - k_{12} + \frac{k_{12}}{a_{12}} \left( k_t^{-1} + a_{12}^{-1} \right)
\right] +  (1,2 \leftrightarrow 3,4) \\ \nonumber
- \frac{k_{12}}{a_{12} a_{34} k_t} \left[ b_{12} b_{34} + 2 k_{12}^2 (\prod_a k_a) \left( 
\frac{1}{k_t^2} + \frac{1}{a_{12}a_{34}} + \frac{k_{12}}{k_t a_{12} a_{34}}
\right) \right],
\end{align}
where $a_{ab} \equiv k_a + k_b + k_{ab}$ and $b_{ab} \equiv (k_a + k_b)k_{ab} + k_a k_b $.

The presence of an overall factor $k_1^2 = q_u^2$ in each of the terms in (\ref{eq:GEcrude}) is very convenient, allowing us to divide the entire expression by $q_u^2$ and then evaluate the remaining part in the limit $\v{q}_u \to 0$. The result will in general depend on the direction from which $\v{q}_u = q_u \hvec{q}_u$ approaches zero. \\

Let's assume that $\hvec{q}_0$ is fixed as $q_u \to 0$. As explained in Section \ref{sec:SqueezedLimits}, we need to take the the sum of the GE contribution over any three\footnote{The final answer would better not depend on which three directions we choose! Otherwise we cannot avoid a major inconsistency.} mutually orthogonal directions of $\hvec{q}_u$. Since $\lim\limits_{q_u \to 0} \left( \mathcal{I}_{abcd} + \mathcal{I}_{cdab} \right)$ does not depend on $\hvec{q}_u$ and the same holds true for $k_{1i}^2$, $\hvec{k}_3 \cdot \hvec{k}_{12}$, $\hvec{k}_2 \cdot \hvec{k}_{13}$ and $\hvec{k}_2 \cdot \hvec{k}_{14}$ (since in each of these instances we can use $\v{k}_{1i} \to \v{k}_i$), all the dependence on $\hvec{q}_u$ is in the following factor:
\begin{equation}
\lim\limits_{q_u \to 0} \left[ \left(1 - (\hvec{q}_u \cdot \hvec{k}_{12} )^2 \right)  \cos 2 \chi_{12,34} \right]
\end{equation}
and its permutations. This is equal to
\begin{equation}
\left(1 - (\hvec{q}_u \cdot \hvec{k}_{2} )^2 \right) \cos 2 \chi_{12,34} \,,
\label{eq:ThisVanishes}
\end{equation}
which can be easily shown to vanish when averaged over any three orthogonal directions of $\v{q}_u$. The same applies to the other two terms in the correlation function. \\

In conclusion, the contribution of the graviton exchange to the $\O (q_u^2)$ term vanishes completely due to averaging over angles.

\subsection{The contact interaction}
The contribution from the contact (4-vertex) interaction reads
\begin{equation}
\label{eq:CIcrude}
\langle \zeta_{\v{k}_1} \zeta_{\v{k}_2} \zeta_{\v{k}_3} \zeta_{\v{k}_4} \rangle'_{CI} = \frac{H_*^6}{4 \epsilon^2 \prod_a (2 k_a^3)} \sum\limits_{\text{24 perms}} \mathcal{M}_4(\v{k}_1, \v{k}_2, \v{k}_3, \v{k}_4).
\end{equation}
where
\begin{align*}
\mathcal{M}_4(\v{k}_1, \v{k}_2, \v{k}_3, \v{k}_4) = - 2 \frac{k_1^2 k_3^2}{k_{12}^2 k_{34}^2} \frac{W_{24}}{k_t} \left( - (\v{k}_1 \cdot \v{k}_4)(\v{k}_2 \cdot \v{k}_3) + (\v{k}_1 \cdot \v{k}_3)(\v{k}_2 \cdot \v{k}_4) + \frac{3}{4} \sigma_{12} \sigma_{34} \right) \\
- \frac{1}{2} \frac{k_3^2}{k_{34}^2} \sigma_{34} \left( \frac{\v{k}_1 \cdot \v{k}_2}{k_t} W_{124} + 2 \frac{k_1^2 k_2^2}{k_t^3} + 6 \frac{k_1^2 k_2^2 k_4}{k_t^4} \right)
\end{align*}
and
\begin{eqnarray*}
\sigma_{ab} & = & \v{k}_a \cdot \v{k}_b + k_b^2, \\
W_{ab} & = & 1 + \frac{k_a + k_b}{k_t} + \frac{2 k_a k_b}{k_t^2}, \\
W_{abc} & = & 1 + \frac{k_a + k_b + k_c}{k_t} + \frac{2(k_a k_b + k_b k_c + k_c k_a)}{k_t^2} + \frac{6 k_a k_b k_c}{k_t^3}. 
\end{eqnarray*}
Using $P(k) = \frac{H_*^2}{4 \epsilon k^3}$, (\ref{eq:CIcrude}) can be written as
\begin{equation}
\langle\zeta_{\v{k}_1} \zeta_{\v{k}_2} \zeta_{\v{k}_3} \zeta_{\v{k}_4}  \rangle'_{CI} =
\epsilon P(k_1) P(k_2) P(k_3) k_4^{-3} \sum\limits_{\text{24 perms}} \mathcal{M}_4(\v{k}_1, \v{k}_2, \v{k}_3, \v{k}_4).
\end{equation}
Our goal is to compute $\frac{1}{2}\frac{\partial^2}{\partial q_u^2} P(q_u)^{-1} \langle \zeta_{\v{q}_u} \zeta_{\v{q}_l} \zeta_{\v{k}_s} \zeta_{\v{k}'_s} \rangle'_{CI}$, so $P(q_u)$ is cancelled out, and the only factor dependent on $\v{q}_u$ is now $\sum\limits_{\text{24 perms}} \mathcal{M}_4(\v{k}_1, \v{k}_2, \v{k}_3, \v{k}_4)$. We have thus reduced the problem to finding
\begin{equation}
\left( \frac{\partial^2}{\partial q_u^2} \sum\limits_{\text{24 perms}} \mathcal{M}_4(\v{k}_1, \v{k}_2, \v{k}_3, \v{k}_4) \right)_{q_u = 0}.
\end{equation}
We will find it easier to deal with the $\mathcal{M}_4(\v{k}_1, \v{k}_2, \v{k}_3, \v{k}_4) $ permutation only, and instead consider all the $24$ different bijections between $(\v{q}_u, \v{q}_l, \v{k}_s, \v{k}'_s)$ and $(\v{k}_1, \v{k}_2, \v{k}_3, \v{k}_4) $.

\subsubsection*{Terms $\O (q_u^0)$, $\O (q_u^1)$}
Using symbolic manipulation in Mathematica, we verified that $O(q_u^0)$ and $O(q_u^1)$ terms in $\mathcal{M}_4$ vanish.

\subsubsection*{The contribution from $\v{q}_u \equiv \v{k}_1$}
Let us first consider the six terms in which $\v{q}_u$ is the first entry. Note that in the limit $q_u \to 0$ we can use $\v{k}_2 + \v{k}_3 + \v{k}_4 = 0$.
\begin{align} \nonumber
\frac{\partial}{\partial q_u^2} \sum\limits_{\text{6 perms}} \mathcal{M}_4(\v{q}_u, \v{k}_2, \v{k}_3, \v{k}_4)  = \sum\limits_{\text{6 perms}} \left[
-2 \frac{k_3^2}{k_2^2 k_{34}^2} \frac{W_{24}}{k_t} \cdot \frac{3}{4} k_2^2 (\v{k}_3 + \v{k}_4) \cdot \v{k}_4 \right. \\ \nonumber
\left.  -  \frac{1}{2} \frac{k_3^2}{k_{34}^2} (\v{k}_3 + \v{k}_4) \cdot \v{k}_4  \left( \frac{\partial}{\partial q_u^2} (\frac{\v{q}_u \cdot \v{k}_2}{k_t} W_{q_u,24})_{q_u=0}  + 2 \frac{k_2^2}{k_t^3} (1+3\frac{k_4}{k_t}) \right)
 \right] \\
 = 
- \sum\limits_{\text{6 perms}}  \frac{k_3^2}{k_{34}^2} (\v{k}_3 + \v{k}_4) \cdot \v{k}_4 \left[
\frac{3}{2} \frac{W_{24}}{k_t} + \frac{k_2^2}{k_t^3} (1+3\frac{k_4}{k_t}) + \frac{1}{4} \left( \frac{\partial^2}{\partial q_u^2} (\frac{\v{q}_u \cdot \v{k}_2}{k_t} W_{q_u,24})  \right)_{q_u=0}
 \right].
\end{align}
Now, we have $\frac{\partial}{\partial q_u}(W_{q_u,24}/k_t)_{q_u = 0} = 0$, so the last term vanishes. We get
\begin{equation}
\frac{\partial}{\partial q_u^2} \sum\limits_{\text{6 perms}} \mathcal{M}_4(\v{q}_u, \v{k}_2, \v{k}_3, \v{k}_4)  = \sum\limits_{\text{6 perms}} \frac{k_3^2}{k_{34}^2} \v{k}_2 \cdot \v{k}_4 \left[
\frac{3}{2} \frac{W_{24}}{k_t} + \frac{k_2^2}{k_t^3} \left( 1+3\frac{k_4}{k_t} \right) 
 \right].
\end{equation}
After some more transformations,
\begin{equation}
\frac{\partial}{\partial q_u^2} \sum\limits_{\text{6 perms}} \mathcal{M}_4(\v{q}_u, \v{k}_2, \v{k}_3, \v{k}_4)  = \sum\limits_{\text{6 perms}} \frac{k_3^2}{k_t^3} \v{k}_2 \cdot \v{k}_4 \left[
1+3\frac{k_4}{k_t} + 3 k_2^{-2} (k_t^2 - \frac{1}{2}k_3 k_t) \right].
\end{equation}
Let's evaluate the dominant term of the above expression for $q_l \ll k_s \sim k'_s$. If it is nonvanishing in the limit $q_l/k_s \to 0$ (we will shortly see that it is), then due to the presence of $q_l$ in the prefactor, the $1+3\frac{k_4}{k_t}$ part gives a zero contribution in this limit. The other term survives as $q_l/k_s \to 0$ only if $\v{q}_l \equiv \v{k}_2$. We are therefore left with
\begin{equation}
 3 q_l^{-2} k_t^{-2} \left( k_s^2 ( \v{q}_l \cdot \v{k}'_s) (k_t - \frac{1}{2}k_s) +  k^{'2}_s \v{q}_l \cdot \v{k}_s (k_t - \frac{1}{2}k'_s) \right) = - \frac{3}{8} k_s (\hvec{k}_s \cdot \hvec{q}_l)^2 + \O (q_l).
\end{equation}

\subsubsection*{The contribution from $\v{q}_u \equiv \v{k}_3$}
Derivation is analogous to that of the previous section, only much simpler. The contribution is
\begin{equation}
\frac{3}{8} k_s [5 (\hvec{k}_s \cdot \hvec{q}_l)^2 - 7].
\end{equation}

\subsubsection*{The contribution from $\v{q}_u \equiv \v{k}_2$}
Using symbolic manipulation in Mathematica, we found
\begin{equation}
\frac{\partial^2}{\partial q_u^2} \mathcal{M}_4(\v{q}, \v{q}_u, \v{k}, \v{K})_{ q_u=0} = 
\frac{k_s^2}{q_l^2 k_t^4} \v{q}_l \cdot \v{k}'_s \left[ 3 k_t^2 (k_t + k'_s) + 2q_l^2 ( k_t + 3k'_s ) \right].
\end{equation}
The only other permutation giving a comparable contribution is the one in which $\v{k}_s$ and $\v{k}'_s$ are swapped. The remaining $4$ permutations of $q_l, k_s, k'_s$ lead to subdominant contributions, which we can ignore. We have
\begin{align} \nonumber
q_l^{-2} k_t^{-4} \left( k_s^2 ( \v{q}_l \cdot \v{k}'_s ) 3 k_t^2 (k_t + k'_s) + 2q_l^2 ( k_t + 3k'_s ) \right. \\
 \left. + k_s^2 \v{q}_l \cdot \v{k}_s 3 k_t^2 (k_t + k_s) + 2q_l^2 ( k_t + 3 k_s ) \right) = \frac{3}{4} k_s (5 (\hvec{k}_s \cdot \hvec{q}_l)^2 - 3) + O(q_l).
\end{align}

\subsubsection*{The contribution from $\v{q}_u \equiv \v{k}_4$}
Again, using symbolic manipulation in Mathematica, we found
\begin{equation}
\frac{\partial^2}{\partial  q_u^2} \mathcal{M}_4(\v{q}_l, \v{k}_s, \v{k}'_s, \v{q}_u)_{ q_u=0} = 
3 \frac{q_l^2}{k^{'2}_s k_t^2} (k_t + k_s) \v{k}_s \cdot \v{k}'_s.
\end{equation}
The dominant contributions when $\v{q}_u \equiv \v{k}_4$ will actually arise from another two permutations: $(k_s,k'_s,q_l)$ and $(k'_s,k_s,q_l)$. We have 
\begin{equation}
3 k_t^{-2} q_l^{-2} ( k^{'2}_s (k_t + k_s) \v{k}_s \cdot \v{q}_l + k_s^2 (k_t + k'_s) \v{k}'_s \cdot \v{q}_l ) =  \frac{3}{4} k_s (5 (\hvec{k}_s \cdot \hvec{q}_l)^2 - 3) + O(q_l).
\end{equation}

\subsubsection*{Summary}

After summing up all the permutations, we get
\begin{equation}
\left( \frac{\partial}{\partial  q_u^2} \sum\limits_{\text{24 perms}} \mathcal{M}_4(\v{k}_1, \v{k}_2, \v{k}_3, \v{k}_4) \right)_{ q_u = 0} \sim \frac{3}{8} k_s (14 (\hvec{k}_s \cdot \hvec{q}_l)^2 - 13)
\end{equation}
for $q_l \ll k_s, k'_s$. 

\subsection{Result - the trispectrum contribution}
Using (\ref{eq:BCA}), we get
\begin{equation}
\boxed{
\langle \zeta_{\v{q}_l} \zeta_{\v{k}_s} \zeta_{-\v{k}_s} \rangle_{K} \sim P(q_l)P(k_s) \left[ (1-n_s)  + \O (q_l/k_s) + \frac{27}{16}  \epsilon \frac{K}{k_s^2} \left( 14 (\hvec{k}_s \cdot \hvec{q}_l)^2 - 13 \right) \right].}
\label{eq:CanBispectrum}
\end{equation}


\section{EFT Trispectrum in the soft limit} \label{app:EFTTrispectrum}


Following \cite{Chen:2009se}, we define the trispectrum form factor $\mathcal{T}$ as
\begin{equation}
\langle \zeta^4 \rangle = 8 \times (2 \pi)^3 \delta^{(3)} \left( \sum\limits_{i=1}^4 \v{k}_i \right) P_{\zeta}(k_1) P_{\zeta}(k_2) P_{\zeta}(k_3) \frac{1}{k_4^3} \mathcal{T}(\v{k}_1, \v{k}_2, \v{k}_3, \v{k}_4)\,,
\label{eq:FormFactorDef}
\end{equation}
where $P_{\zeta}(k) = \langle \zeta_{\v{k}} \zeta_{-\v{k}} \rangle'$. We will use the results of \cite{Chen:2009se}; the parameters $\lambda$ and $\Sigma$ from \cite{Chen:2009se} are related to $c_s$ and $c_3$ in the following way:
\begin{eqnarray}
\lambda & = & \frac{1}{2} \Sigma  \left( 1 + \frac{2}{3} \frac{c_3}{c_s^2} \right)  , \\
\Sigma & = & \frac{\epsilon H^2}{c_s^2}.
\end{eqnarray}
Throughout, we assume that $c_s \ll 1$, while $c_3 \sim \O (1)$, so that $\lambda \sim \O (c_s^{-4})$, $\Sigma \sim \O (c_s^{-2})$.

\subsection{The EFT power spectrum}
For models with small speed of sound, we have
\begin{equation}
P_{\zeta}(k) = \frac{H_*^2}{4 \epsilon M_{pl}^2 c_s k^3}.
\end{equation}

\subsection{scalar-exchange diagram}
The scalar exchange contributes to the trispectrum at order $c_s^{-4} q_u^2 k_s P_{\zeta}^4$. In particular, there are no terms at $0^{th}$ or $1^{st}$ order in the ultralong momentum $q_u$, as has been verified by our Mathematica scripts. In the computations we outline in this subsection, all contributions that are subleading in the regime $q_u \ll k_l \ll k_s$ are neglected.

The dominant contributions come from the diagrams in which the exchanged momentum is $\v{q}_u + \v{k}_l$, corresponding to the choices $(\v{q}_u, \v{k}_l) \equiv (\v{k}_3, \v{k}_4)$ or $(\v{q}_u, \v{k}_l) \equiv (\v{k}_1, \v{k}_2)$, where $(a,b)$ stands for an unordered pair. Hence, we only have to sum over 8, rather than 24, permutations of the momenta.

We also perform a summation over three\footnote{Of course, the $O(q_u^2)$ terms are even in $\v{k}_u$, so we do not need to average over two opposite directions.} directions of $\v{q}_u$ that are mutually orthogonal, but otherwise arbitrary.

There are three types of scalar-exchange diagrams that differ by the type of vertices, shown on Fig. \ref{feynmans}. For the sake of transparency, we consider each of the three cases in a separate subsection, writing out the partial contributions before presenting the final result.

\subsubsection*{The $\dot{\pi}^3 \times \dot{\pi}^3$ diagrams}
These diagrams are given by (B.3) - (B.4) in \cite{Chen:2009se}. It is straighforward to show that the leading-order contribution to $\mathcal{T}$ is
\begin{equation}
\mathcal{T}_1 = \frac{9}{32} \left( \frac{\lambda}{\Sigma} \right)^2 q_u^2 k_s = \frac{9}{128} \left(1 +  \frac{2}{3} \frac{c_3}{c_s^2} \right)^2 q_u^2 k_s.
\end{equation}
After summing over three orthogonal directions of $\v{q}_u$,
\begin{equation}
\sum\limits_{\hvec{q}_u} \mathcal{T}_1 = \frac{27}{128} \left(1 +  \frac{2}{3} \frac{c_3}{c_s^2} \right)^2 q_u^2 k_s.
\end{equation}

\subsubsection*{The $\dot{\pi}^3 \times \dot{\pi}(\partial_i \pi)^2$ and $\dot{\pi}(\partial_i \pi)^2 \times \dot{\pi}^3$ diagrams}
These are given by (B.5) - (B.10) in \cite{Chen:2009se}. By computing $\partial^2/\partial q_u^2$ of the sum of the eight relevant permutations, we can deduce the $q_u^2$ term. We find, to leading order in $k_s/k_l$,
\begin{equation}
\sum\limits_{\hvec{q}_u} \mathcal{T}_2 = \frac{3}{128} \left( 1 + \frac{2}{3} \frac{c_3}{c_s^2} \right) \left( \frac{1}{c_s^2}-1 \right) q_u^2 k_s \left( 43 - 15 ( \hvec{k}_l \cdot \hvec{k}_s )^2 \right).
\end{equation}

\subsubsection*{The $\dot{\pi}(\partial_i \pi)^2 \times \dot{\pi}(\partial_i \pi)^2$ diagrams}

These diagrams correspond to equations (B.11) - (B.17) from \cite{Chen:2009se}. Again, after summing over the eight permutations, computing the $\partial^2/\partial q_u^2$ derivative, summing over the three directions and keeping only the terms that are leading order in $k_s/k_l$, we get
\begin{equation}
\sum\limits_{\hvec{q}_u} \mathcal{T}_3 = \frac{19}{128} \left( \frac{1}{c_s^2}-1 \right)^2  q_u^2 k_s \left( 8 - 5 ( \hvec{k}_l \cdot \hvec{k}_s )^2 \right).
\end{equation}

\subsubsection*{Summary for the scalar exchange}
The total form factor due to scalar exchange is, after summing over angles,\footnote{The factor of $1/8$ is introduced in order to cancel out the factor of $8$ in front of $\mathcal{T}$ in \eqref{eq:FormFactorDef}.}
\begin{equation}
\mathcal{T}_{SE} = \frac{1}{8} q_u^2 k_s \left( \alpha_1 + \alpha_2 ( \hvec{k}_l \cdot \hvec{k}_s )^2 \right),
\end{equation}
with $\alpha_1$ and $\alpha_2$ that can be expressed in terms of $c_s, c_3$. If we keep only the $c_s^{-4}$ terms, we have
\begin{eqnarray}
\alpha_1 & = & \left( \frac{3}{4} c_3^2 + \frac{43}{8}c_3 + \frac{19}{2} \right) c_s^{-4}, \\
\alpha_2  & = & - \left(\frac{15}{8}c_3 + \frac{95}{16} \right) c_s^{-4}.
\end{eqnarray}
Then
\begin{equation}
\sum\limits_{\hvec{q}_u} \langle \zeta^4 \rangle_{SE} \sim (2 \pi)^3 \delta^{(3)} \left( \sum\limits_{i=1}^4 \v{k}_i \right) P_{\zeta}(q_u) P_{\zeta}(k_l) P_{\zeta}(k_s) \frac{q_u^2}{k_s^2}\left( \alpha_1 + \alpha_2 ( \hvec{k}_l \cdot \hvec{k}_s )^2 \right).
\end{equation}
The corresponding contribution to the bispectrum on a curved background is
\begin{equation}
\langle \zeta_{\v{k}_l} \zeta_{\v{k}_s} \zeta_{-\v{k}_s} \rangle'_{K,SE} \sim  \frac{3}{2}  P_{\zeta}(k_l) P_{\zeta}(k_s)  \left( \alpha_1 + \alpha_2 ( \hvec{k}_l \cdot \hvec{k}_s )^2 \right)  \frac{K}{k_s^2} .
\label{eq:BipectrumApp}
\end{equation}
The above power spectra are evaluated on a flat background. But $\alpha_i$ scale as $c_s^{-4}$ while the correction to the power spectrum scales as $c_s^{-2}$, so it might be neglected in the regime $c_s \ll 1$. Then in (\ref{eq:BipectrumApp}) we are allowed to use the curved power spectrum.

\subsection{Graviton exchange}
Another tree-level contribution to the scalar $4-$pt function is the graviton exchange. However, as we have shown in Section \ref{sec:SqueezedLimits}, the contribution of the lowest-order operators to the final $O(K)$ terms in the bispectrum is always exactly zero.

\subsection{Contact interaction}\label{contactint}
There is yet another contribution to the trispectrum that cannot be accounted for in the cubic action, since it originates from the contact diagram corresponding to the $4-$vertex scalar interaction. The contact diagram is evaluated in \cite{Chen:2009se}. In the limit $c_s \to 0$, the result is dominated by
\begin{equation}
\mathcal{T}_{c1} = 36 \left(C_{\dot{\pi}^4}- 9 \left( \frac{\lambda}{\Sigma} \right)^2 \right) \frac{\prod_{i=1}^{4} k_i^2}{k_t^5}.
\end{equation}
In the double squeezed limit, this gives
\begin{equation}
\langle \zeta^4 \rangle\,_{c1} \propto  \left( C_{\dot{\pi}^4} - 9 \left( \frac{\lambda}{\Sigma} \right)^2 \right) P_{\zeta}(q_u) P_{\zeta}(k_l) P_{\zeta}(k_s) \frac{q_u^2 k_l^2}{k_s^4}.
\end{equation}
Recall that $\left( \lambda / \Sigma \right)^2 \sim \O (c_{s}^{-4})$. For the particular case of DBI inflation, $C_{\dot{\pi}^4}$ also scales as $c_s^{-4}$; we assume, for simplicity, that the scaling of the above contribution is always $c_s^{-4}$. There are also other terms that scale as $c_s^{-2}$ and are subdominant in the limit $c_s \to 0$. These are the only contributions to the trispectrum that are linear in $q_u$ for small $q_u$ (it has been verified in \cite{Creminelli:2012ed} that these terms reproduce the conformal consistency relation for the $4-$pt function). 

Summing up everything, we have (schematically)
\begin{equation}
\langle \zeta^4 \rangle\,_{c} \sim P_{\zeta}^3 \left[c_s^{-2} \left( \frac{q_u k_l}{k_s^2} + \frac{q_u^2}{k_s^2} \right) + c_s^{-4} \frac{q_u^2 k_l^2}{k_s^4} \right].
\end{equation}
Due to the nonvanishing $O(q_u)$ terms, redefinitions of the momenta will influence the $O(q_u^2)$ terms; but only at order $c_s^{-2}$, not $c_s^{-4}$: 
\begin{equation}
\langle \zeta^4 \rangle\,^{(q_u^2)}_{c} \sim P_{\zeta}^3 \left[c_s^{-2} \frac{q_u^2}{k_s^2} + c_s^{-4} \frac{q_u^2 k_l^2}{k_s^4} \right].
\end{equation}
The first term in the brackets is subdominant in the limit $c_s \to 0$ relative to the scalar-exchange diagram. The second term is also subdominant, having a different momentum dependence than the leading-order scalar exchange contribution. 

In conclusion, the contact interaction gives a negligible contribution to the $\O (K)$ correction to the squeezed bispectrum. The final result for the squeezed limit of the bispectrum in a curved universe is given by (\ref{eq:BipectrumApp}).

\bibliographystyle{utphys}
\bibliography{refs}


\end{document}